\documentclass[11pt]{article}

\makeatletter
\@addtoreset{equation}{section}
\makeatother

\usepackage{epsfig,amsfonts}
\usepackage[fleqn]{amsmath}
\usepackage{amsthm,amssymb}
\usepackage{yfonts,mathrsfs}
\usepackage{graphicx}
\usepackage{hhline}
\usepackage{cite}
\usepackage{upgreek}
\def\be{\begin{equation}}
\def\ee{\end{equation}}
\def\bdm{\begin{displaymath}}
\def\edm{\end{displaymath}}
\def\bea{\begin{eqnarray}}
\def\eea{\end{eqnarray}}

\def\Xint#1{\mathchoice
    {\XXint\displaystyle\textstyle{#1}}%
    {\XXint\textstyle\scriptstyle{#1}}%
    {\XXint\scriptstyle\scriptscriptstyle{#1}}%
    {\XXint\scriptscriptstyle\scriptscriptstyle{#1}}%
    \!\int}
\def\XXint#1#2#3{{\setbox0=\hbox{$#1{#2#3}{\int}$}
    \vcenter{\hbox{$#2#3$}}\kern-.5\wd0}}

\def\dashint{\Xint-}

\begin{document}
%\draft
%%%%%%%%%%%%%%%%%%% Title %%%%%%%%%%%%%%%%%%%%%%%%%%%%%%%%%%%%%%%%%%%
\begin{titlepage}

\begin{flushright}
RUNHETC-2006-13
\end{flushright}

\vspace{0.2cm}

\begin{center}

\begin{LARGE}

{\bf Ising Spectroscopy I: Mesons at $T < T_c$}

\end{LARGE}

\vspace{0.6cm}

\begin{large}

{\bf P. Fonseca and  A. Zamolodchikov$^{1,2}$}

\end{large}

\vspace{0.3cm}

{${}^{1}$NHETC, Department of Physics and Astronomy\\
     Rutgers University\\
     Piscataway, NJ 08855-0849, USA\\

\vspace{0.2cm}

${}^{2}$L.D. Landau Institute for Theoretical Physics\\
  Chernogolovka, 142432, Russia
}

\vspace{0.3cm}

\end{center}

\begin{center}
\centerline{\bf Abstract} 
\vspace{0.3cm}
\parbox{12.5cm}{
This paper is our progress report on the project ``Ising
spectroscopy'', devoted to a systematic study of the mass spectrum of
particles in the 2D Ising Field Theory in a magnetic field. Here we address 
the low-temperature regime, and develop a quantitative approach based on 
the idea (originally due to McCoy and Wu) of particles being the
``mesons'', consisting predominantly of two quarks confined by a long-range
force. Systematic implementation of this idea leads to a version of
the Bethe-Salpeter equation, which yields infinite sequence of meson
masses. The Bethe-Salpeter spectrum becomes exact in the limit when
the magnetic field is small, and we develop the corresponding
weak-coupling expansions of the meson masses. The Bethe-Salpeter
equation ignores the contributions from the multi-quark components of
the meson's states, but we discuss how it can be improved by treating
these components perturbatively, and in particular by
incorporating the radiative corrections to the quark mass and the
coupling parameter (the ``string tension''). The approach fails to
properly treat the mesons above the stability threshold, where they
are expected to become resonance states, but it is shown to yield a very
good approximation for the masses of all stable particles, at all
real values of the IFT parameters in the low-temperature regime. We
briefly discuss how the Bethe-Salpeter approximation can be used to
address the case of complex parameters, which was the main motivation
of this work.
}

 \vspace{0.4cm}

\begin{flushleft}

\rule{4.0 in}{.007 in}\\
{\ \ December 2006}
\end{flushleft}
\vfill

\end{center}

\end{titlepage}
\newpage

\section{ Introduction} \label{intro}

The Ising Field Theory (IFT) is the quantum field theory in two
dimensions which emerges in the scaling limit of the Ising model (or
any other system from the same universality class) near its critical
point. The RG fixed point associated with the Ising criticality is
the $c=1/2$ Minimal Conformal Field Theory (see e.g. \cite{drouffe,cft}) which
has two nontrivial relevant operators. Therefore the IFT is understood
as the Minimal CFT perturbed by the two relevant operators, as
described by the Euclidean Action
\bea\label{ift}
{\cal A}_{\rm IFT} = {\cal A}_{\rm c={1/2}\ CFT} +
\tau\,\int\,\varepsilon (x)\,d^2 x + h\,\int\,\sigma(x)\,d^2 x\ ,
\eea
where the fields $\varepsilon(x)$ (``energy density'') and
$\sigma(x)$ (``spin density'') have conformal dimensions
$({1\over 2},{1\over 2})$ and $({1\over 16},{1\over 16})$,
respectively$\,$\footnote{Here we assume conventional normalization of the 
fields, such that
$$
\langle\,\,\varepsilon(x)\,\varepsilon(0)\,\rangle \to
|x|^{-2} \qquad {\rm and} \qquad \langle\,\sigma(x)\sigma(0)\,\rangle \to
|x|^{-{1\over 4}} \qquad {\rm as}\quad x \to 0\ .
$$

This condition fixes the normalizations of the couplings $\tau$ and
$h$.}. The parameters $\tau$ and $h$ are scaled deviations of the
temperature $T$ and the external magnetic field $H$ of the Ising 
model from the critical values, i.e. $\tau \simeq T_c-T$
and $h \simeq H$ in the limit $T_c-T\to 0, H\to 0$. In what
follows we refer to $h$ as the magnetic field. It is assumed here
that the definition of $\varepsilon(x)$ is such that the domain
$\tau >0$ corresponds to the low-temperature (ordered) phase
of the model. The parameters $\tau$ and $h$ are dimensionfull,
$\tau \sim [length]^{-1}$ and $h \sim [length]^{-{15\over
    8}}$. Therefore the physics of IFT essentially depends on a
single dimensionless scaling parameter, which we define 
as\,\footnote{This is the same definition of $\eta$ as was used in 
\cite{fz1,fz2}\,; note that it differs from definition 

in Refs.\cite{delfino1,caselle,res1}.}

\bea\label{etadef}
\eta\ =\ \frac{m}{\ \,|h\,|^{\frac{8}{15}}} \ \,\equiv\
\,2\pi\,\frac{\tau}{\ |h\,|^{\frac{8}{15}}}\ ,
\eea
where we have also introduced the notation  $m = 2\pi\,\tau$ which is
extensively used below.

Besides its value as the model of near-critical statistical mechanics,
the IFT appears to be rich a model of two-dimensional particle theory.
It was argued in the pioneering work of McCoy and Wu \cite{mccoy2}
that as one changes $\eta$ from $-\infty$ to $+\infty$,
the spectrum of stable particles undergoes evolution from a single
particle to an infinite tower of ``mesons'' formed by weekly confined
``quarks''.  This scenario was subsequently confirmed by exact
\cite{me1} and numerical \cite{delfino2,fz1}\ results, where some
quantitative details were added.

The IFT admits an exact solution in few special cases. First, at zero
magnetic field, i.e. at $\eta = \pm\,\infty$, the IFT is well-known
to be the theory of free Majorana fermions (see e.g. 
\cite{mccoy1,drouffe}) 
of mass $|m|$. Second, the IFT is shown to be integrable
at $m=0$ and $h\neq 0$ (i.e. at $\eta=0$) \cite{me1}. In the last
case, the associated particle theory involves eight stable particles 
whose scatterings are described by certain factorizable S-matrix. 
Finally, if one admits complex values of the parameters in
\eqref{ift}, yet another integrable theory is attributed to the
vicinity of the point(s) where $h$ takes special pure-imaginary value
$h = \pm i\,(0.1893...)\,m^{15\over 8}$. This point is the Yang-Lee
edge singularity \cite{yl}. In the vicinity of this point one of the
particles of the theory becomes light, and its dynamics is described
by the integrable Yang-Lee field theory \cite{fisher,cardy1,cardy2}. 
At present there are no indication that IFT \eqref{ift} might be
integrable at any other values of the parameters, real or complex.

This paper is our progress report on (a part of) the program ``Ising 
spectroscopy''. The program is about the detailed study of the mass
spectrum of the IFT at all values of the scaling parameter (with the
emphasis on the analytic properties of the masses as the functions of
$\eta$), using perturbative
expansions around the integrable points, combined with numerical
methods. There are several motivations for investing effort in this 
program. For one, the IFT appears in some ways the most basic unitary 
quantum field theory in two dimensions. The Ising fixed point is the 
lowest (in terms of the central charge) nontrivial unitary CFT in 2D 
\cite{cft}, and the theory \eqref{ift} corresponds to generic RG flow
which originates at this point. Also, the IFT is a playground for 
studying many phenomena common in QFT, like quark confinement, resonance
states, first order phase transition and associated ``false'' vacuum 
state, etc.

In this paper we discuss the mass spectrum of the IFT at real $h$ and
$m$, in the low-temperature (low-T) domain $m>0$, i.e. at real $\eta >
0$. Let us remark here that at $h\neq 0$ distinction between the low-T
and high-T ($\eta <0$) regimes is fuzzy, and to a large extent
artificial. Physical quantities like the particle masses pass through
the point $\eta=0$ continuously (in fact, analytically). However,
although continuously related, the two regimes call for
an emphasis on different phenomena, and often require different
theoretical treatment. Therefore the distinction is conventional
in the literature on the subject, and we will usually hold to it here
and in forthcoming reports on our program.

According to McCoy-Wu scenario, at $\eta < \infty$ the free
fermions of the $h=0$ theory (the ``quarks'') become confined in
``mesons''. The spectrum of the mesons is dense at large $\eta$.
As $\eta$ decreases, the heavier mesons become unstable against decay
into the lighter ones, and successively disappear from the spectrum
of stable particles, typically becoming resonances. As $\eta$
approaches zero, there are only three stable mesons (and many
resonances) left (see \cite{fz1}), and at $\eta=0$ some of the
resonances regain stability (the phenomenon related to integrability
of the IFT at $\eta=0$), so there are exactly eight stable particles
present\,\footnote{The process of
depletion of the spectrum continues when $\eta$ crosses to the
negative values, until below a certain point ($\eta < \eta_2 \approx
-2.09$, see \cite{fz1}) there is only one stable particle left. At $\eta \to
-\infty$ all interactions disappear and this particle becomes a free
one, in accord with the IFT at zero $h$ 
being a theory of free fermions. We have much more to say about
detailed structure of the spectrum in this domain, but will deffer
this discussion to another paper.}. Very accurate numerical data for 
the particle masses $M_n$
at all $\eta$ can be obtained using the ``Truncated Free Fermion Space
Approach'' (TFFSA)\,\footnote{The TFFSA is a modification of
the ``Truncated Conformal Space Approach'' (TCSA) of Ref.\cite{tcsa}, 
particularly suited for the IFT, see Ref.\cite{fz1} for details.}; the 
results for positive $\eta$ are shown in Fig.1. 
\begin{figure}
\centering
\includegraphics[width=13.5cm]{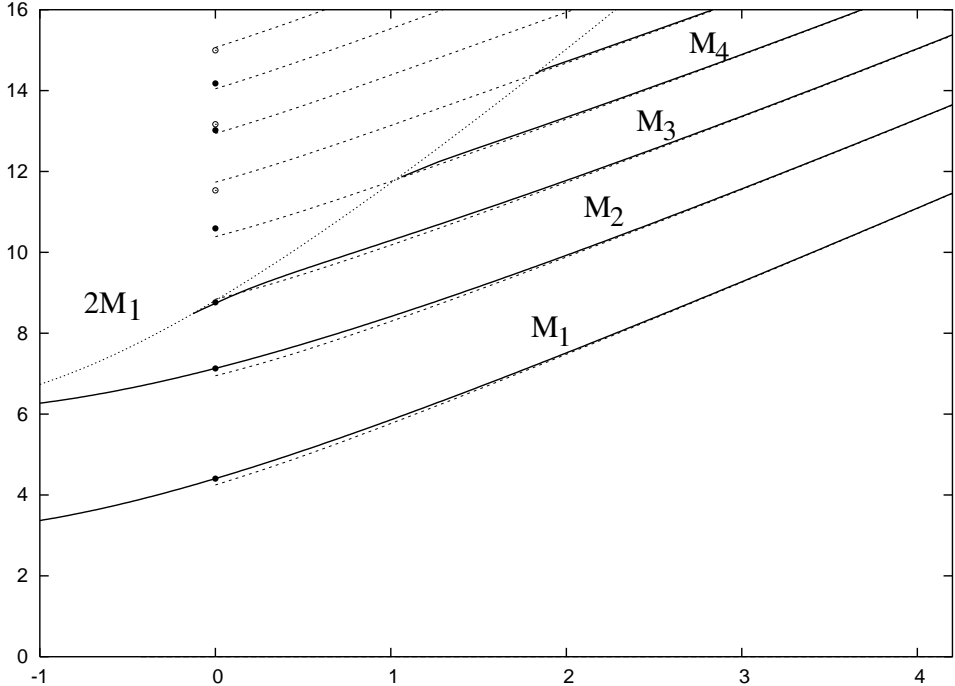}
\caption{\small The mass spectrum $M_n(\eta)$ of IFT particles at positive
(and some negative) $\eta$. The solid lines represent numerical data
  obtained using the TFFSA. The dotted line shows the stability
  threshold $2 M_1$; after crossing this line the particles
  become unstable, and their masses (not shown) develop imaginary parts.
The exception is the point $\eta=0$, where there are eight stable
particles, whose masses are indicated by bullets $\bullet$\,. The
dashed lines show the masses obtained from the Bethe-Salpeter equation
\eqref{bs}, with the renormalized string tension $f$ defined as in
Eq.\eqref{fre} (see Section 8 for explanations). 
The circles $\circ$ indicate positions of the higher thresholds $M_1 +
M_2$, $M_1 + M_3$, etc, at $\eta=0$.}
\label{mass-plot}
\end{figure}
The form of the mass spectrum at sufficiently large $\eta$ is
consistent with the McCoy-Wu idea of the particles being composed of
two weakly confined quarks. This idea allows one to
develop the weak-coupling (large-$\eta$) expansions of the masses $M_n
(\eta)$. The leading term was first obtained in the original paper
\cite{mccoy2}, and several further terms added in Ref.\cite{fz1}. On the
other hand, for highly excited mesons the semiclassical approximation 
applies, and the corresponding expansion was recently derived 
by Rutkevich \cite{rut}\,\footnote{We have worked out the
  semiclassical expansion
independently, before \cite{rut} appeared. Our approach is described
in the Section 5 below. We would like to stress that beyond the
leading order our result \eqref{wkbd} differs from the corresponding
term found in \cite{rut}. The discrepancy can be traced to the 
corrections to the quark self-energy, which we believe are properly
taken into account in the Bethe-Salpeter equation \eqref{bs},
but are missing in the treatment in \cite{rut}.}. 

In this work we develop a more systematic approach based on the
two-quark approximation. In Sections 3 and 4 we derive the
Bethe-Salpeter (BS) equation, i.e. the relativistic two-quark bound
state equation for the IFT mesons. It is not exactly a textbook
exercise because no covariant perturbation theory for the IFT
is available, and one has to do with the ordinary non-covariant
approach. In Section 4 we argue that the Lorentz covariance is
restored by adding certain corrections to the quark self-energy, which
technically come from the multi-quark sectors. The resulting BS
equation (Eq.\eqref{bs}) is very similar to the bound-state equation
previously obtained by 't Hooft for the two-dimensional model of
multicolor QCD \cite{thooft}, which is not surprising because the
physics is similar. The two-quark approximation can be improved by
taking the multi-quark sectors into account perturbatively. In Section
6 we discuss the role of the multi-quark corrections, in particular
the effects of the renormalizations of basic parameters - the quark
mass and  ``string tension'' - due to the multi-quark effects. In
Sections 7 and 8 we find the spectrum of the BS equation numerically, 
and make a tentative comparison with the TFFSA data. If the
renormalization of the string tension is taken into account, the BS
spectrum reproduces the actual masses rather accurately at all
positive $\eta$, as is shown in Fig.1. The actual masses $M_n (\eta)$
exhibit a subtle behavior\footnote{The near-threshold behavior of the
meson masses $M_n (\eta)$ deserves a separate discussion. Preliminary
analysis (which will be reported elsewhere) shows the following
general pattern. The $n$-th meson exists as a stable particle while
$\eta$ exceeds certain threshold value, $\eta > \eta_n$, in which
domain $M_n (\eta)$ remains below $2 M_1 (\eta)$. At $\eta=\eta_n$ the 
graph of $M_n (\eta)$ touches the curve $2 M_1 (\eta)$, so that $2 M_1
(\eta) - M_n (\eta) \sim (\eta_n - \eta)^2$ as $\eta \to
\eta_n$. Correspondingly, the analytic continuation of $M_n
(\eta)$ to $\eta$ immediately below $\eta_n$ remains real, although no
stable particle with this mass exists at $\eta < \eta_n$ (the
situation known as the ``virtual level'', see e.g. \cite{landau}, \S
128). By further decreasing the value of $\eta$ one reaches
singularity of the function $M_n (\eta)$ (typically a square-root
branching point) at $\eta = \hat\eta_{n} < \eta_n$, and  below
$\hat\eta_n$ the mass $M_n(\eta)$ develops an imaginary part, so that the
$n$-th meson reemerges as a resonance state. This pattern seems to
hold for all $M_n$ with $n \geq 3$. Our preliminary estimates of the
thresholds are $\eta_2 \approx -2.06$, $\eta_3 \approx -0.136$,
$\eta_4 \approx 1.0$, while $\hat\eta_3 \approx -0.515$, and the higher
$\hat\eta_n$ are very close to $\eta_n$.} when they approach
the ``stability threshold'' $2 M_1$ (it is visible in Fig.1 for the
mass $M_3$, and less prominently for $M_4$ and $M_5$), the feature
certainly driven by the multi-quark components of the meson
states. The Bethe-Salpeter 
approximation does not capture these subtleties. Also, the BS masses 
remain real above the stability threshold, while the actual IFT
particles in this domain generally become unstable, and their
masses develop imaginary parts. Although in this paper we do not
discuss the resonance part of the IFT spectrum in any detail (we plan
to address this important subject separately; some results on the IFT
resonances can be found Ref's \cite{delfino2,res1}), let us note that
the pattern shown in Fig.1 suggests that when $\eta$ goes to zero some
of the the IFT resonances turn into the particles of the corresponding
integrable theory, while others become weakly coupled bound states of
those, with the binding energy $\sim\eta^2$. For instance, 
\bea\label{mfive}
M_5 (\eta) = M_1 (\eta) + M_2 (\eta) + O(\eta^2) \quad {\rm as} \quad
\eta\to 0\,.
\eea
This is one of many interesting phenomena which the IFT spectrum
appears to exhibit; it provides illustration of its remarkable
richness, and gives us additional motivation for its systematic
study. 

\section { Ising quarks and Ising mesons}\label{qm}

At $h=0$, and in the low-temperature phase $m>0$, the spin-reversal
symmetry of the IFT is spontaneously broken, and the theory has two
degenerate ground states differing by the sign of the spontaneous
magnetization
\bea\label{magnetization}
\langle\,\sigma(x)\,\rangle_{\pm} = \pm \,{\bar\sigma}\ ,
\eea
where ${\bar \sigma}$ is a constant whose exact value is \cite{tracy}
\bea\label{sexact}
{\bar \sigma} = |m|^{1\over 8}\,{\bar s}\ , \ \ \ \ \
{\bar s} = 2^{1/{12}}\,\,e^{-{3\over 2}\,\zeta'(-1)} =
1.35783834170660...
\eea
At the same time, as was mentioned in the Introduction, at $h=0$ the
IFT \eqref{ift} reduces to the theory of free Majorana fermions of the
mass $m$. These free fermions are identified with the domain walls
separating the spatial domains of positive and negative
magnetization. Such interpretation makes it obvious that adding the
interaction term $h\int\sigma(x) d^2 x$ with small $h$
($h<<m^{15\over 8}$) to the free theory generates confining force between
the domain walls, with the ``string tension'' $2{\bar\sigma}h$. Due to
this effect, the stable particles appear as
the ``mesons'', i.e. the bound states of two ``quarks'' \footnote{The
  gauge group associated with this confining interaction is $Z_2$ (the
Ising model with magnetic field is known to be dual to the $Z_2$ gauge
theory, see e.g. \cite{drouffe}), hence the quarks and the antiquarks 
are identical.}.

The above picture alone allows one to give some 
quantitative description of the masses $M_n$, $n=1,2,3,...$
of the mesons in the limit $\eta\to +\infty$. Indeed, the Hamiltonian
of two particles interacting via the confining force is
\bea\label{htwo}
H = \omega(p_1) + \omega(p_2) + 2{\bar\sigma}h\,|x_1-x_2|\ ,
\eea
where $x_1,x_2$ and $p_1,p_2$ are the coordinates and the momenta of
the quarks, and
\bea\label{omegadef}
\omega(p) = \sqrt{m^2 + p^2}\ .
\eea
For the lower part of the meson spectrum ($M_n$ with $n$ fixed)
the non-relativistic approximation
$\omega(p) \approx m + {{p^2}/{2m}}$ applies,
and the problem reduces to finding energy levels of the Hamiltonian
${{p^2}/m} + 2{\bar\sigma}h\,|x|$. Since the quarks are fermions,
only odd levels are relevant, and one finds
\bea\label{mccoys}
M_n - 2m \ \to\ \big(2{\bar\sigma} h\big)^{2\over 3}\,z_n \ \ \ \ {\rm
  as} \ \ \ h \to 0
\eea
where $z_n, n=1,2,...$ are consecutive zeros of the function ${\rm
  Au}(-z)$. This result was first obtained by McCoy and Wu
\cite{mccoy2} in their analysis of the spin-spin correlation
function. For higher mesons ($M_n$ with $n \sim m^2/{\bar\sigma} h$)
the semiclassical approximation is more suitable. Classical
trajectories of the system \eqref{ht} are periodically repeated
cycles in which, during the time between two collisions, the quarks
move under constant acceleration directed towards each other. It is
convenient to parametrize the time $t$ as $(t-t_0) =
R\,\sinh\beta$, where $t_0$ is suitable reference
time, and $R = m/(2{\bar\sigma}h)$.
The parameter $\beta$ has a simple meaning: $+\beta$ and $-\beta$ are
the rapidity of the two quarks in the center-of-mass frame. In this
frame, the classical trajectory within each cycle is given by the equations
(up to permutation of the quarks)
\bea\label{cycle}
p \equiv {{(p_1 - p_2)}\over 2} = - m\,\sinh\beta\ , \ \ \ x \equiv
x_1-x_2 =
2 R\,
\big(\cosh\vartheta - \cosh\beta\big)
\eea
for $-\vartheta < \beta < \vartheta$, where $\vartheta$ is positive
parameter characterizing the classical trajectory. At
$\beta=\pm\,\vartheta$ the separation between the quarks vanishes,
hence the mass (i.e. the center-of-mass energy \eqref{ht})
associated with the trajectory is $M = 2m\,\cosh\vartheta$.
The reduced action per cycle is
\bea\label{pdx}
\int_{\rm cycle} \,p dx =
2mR\,\int_{-\vartheta}^{\vartheta}\,\sinh^2\beta\,d\beta =
{{\sinh 2\vartheta - 2\vartheta}\over{\lambda}}\ ;
\eea
here and below we use the notation $\lambda$ for the dimensionless 
ratio 
\bea\label{lambdadef}
\lambda = {{2{\bar\sigma} h}\over{m^2}}\ .
\eea
If one first treats the quarks as distinguishable particles, the full
period consists of two cycles, so that the reduced action per period
is twice the value of \eqref{pdx}, so that the Bohr-Sommerfeld
quantization condition reads $2\,\int_{\rm cycle}\,pdx = 2\pi
\big(N+{1/2}\big)$, where the integers $N$ must be taken to be
odd, $N=2n-1$, to reclaim the fermionic nature of the quarks. The
quantization condition then takes the form
\bea\label{bscondition}
{{\sinh 2\vartheta_n - 2\vartheta_n}\over{\lambda}} =
2\pi\,\big(n-{1/4}\big)\ ,
\ \ \ \ \ n=1,2,3,\ldots
\eea
leading to the WKB mass spectrum of the mesons,
\bea\label{wkbmass}
M_n = 2m\,\cosh\vartheta_n\ .
\eea

\section {The two-quark approximation}\label{twoq}

The idea of the Ising particles being predominantly two-quark
composites allows one to develop somewhat a more systematic theory of
mesons, in particular to develop a weak-coupling expansions
for their masses $M_n$. One starts with the free-fermion theory at
$h=0$, and introduces the creation and annihilation operators
${\bf a}^{\dagger}_p$ and ${\bf a}_p$ (where $p$ denotes the spatial
momentum of the particle), subject to the canonical anticommutators
$\{{\bf a}_p , {\bf a}^{\dagger}_{q}\} = 2\pi\,\delta(p-q)$.
Below we will use the notation
\bea\label{freestates}
\mid p_1,p_2,\cdots,p_n\,\rangle \ \equiv \ {\bf a}_{p_1}^{\dagger}
{\bf a}_{p_2}^{\dagger}\cdots {\bf a}_{p_n}^{\dagger}\mid 0\,\rangle
\eea
for the $n$-particle states.

The $h=0$ theory is described by the free Hamiltonian
\bea\label{hfree}
{\bf H}_0 = E_0 + \int_{-\infty}^{\infty}\,\omega(p)\,{\bf
  a}^{\dagger}_p {\bf a}_p\,\,{{dp}\over{2\pi}}\ ,
\eea
where $\omega(p)$ is the relativistic energy \eqref{omegadef}\ , and
$E_0$ is the ground-state energy. The
interaction is generated by adding the term associated with the last
term in the action \eqref{ift},
\bea\label{hift}
{\bf H} = {\bf H}_{0} + h\,\int_{-\infty}^{\infty}\,\sigma({\rm
  x})\,d{\rm x}\ ,
\eea
where ${\rm x}$ is the spatial coordinate, and $\sigma({\rm x}) =
\sigma({\rm x},{\rm t})|_{{\rm t}=0}$. The operator $\sigma({\rm x})$
can not be expressed through the fermions in a local way (see
e.g. \cite{drouffe}),
but its matrix elements between the states with any number of the
fermions are known explicitly \cite{karowski}. Here we will only need the
$2\to 2$ matrix element
\bea\label{matrix}
\langle\,p_1,p_2\mid \,\sigma(0)\,\mid q_1, q_2 \,\rangle =
4{\bar\sigma}\,\,{\cal G}(p_1,p_2|q_1,q_2)\ ,
\eea
where
\bea\label{ffact}
{\cal G}(p_1,p_2|q_1,q_2) =
{{(1/4)}\over{\sqrt{\omega(p_1)\omega(p_2)\omega(q_1)\omega(q_2)}}}\,
\bigg[{{\omega(p_1)+\omega(q_1)}\over{p_1-q_1}}\,
{{\omega(p_2)+\omega(q_2)}\over{p_2-q_2}}- \nonumber \\
{{\omega(p_1)+\omega(q_2)}\over{p_1-q_2}}\,
{{\omega(p_2)+\omega(q_1)}\over{p_2-q_1}} +
{{p_1-p_2}\over{\omega(p_1)+\omega(p_2)}}\,{{q_1-q_2}
\over{\omega(q_1)+\omega(q_2)}} \bigg]
\eea
One looks for the eigenstates of the Hamiltonian \eqref{hift}
of the form
\bea\label{mesonstate}
\mid \Psi\,\rangle =\mid \Psi^{(2)}\,\rangle +
\mid \Psi^{(4)}\,\rangle + \mid \Psi^{(6)}\,\rangle + \cdots
\eea
where
\bea\label{psitwo}
\mid \Psi^{(2)}\,\rangle = {1\over
  2}\,\int_{-\infty}^{\infty}\,{{dp_1}\over{2\pi}}\,{{dp_2}\over{2\pi}}\,
\Psi (\,p_1\,,\,p_2\,)\,\mid p_1,p_2\,\rangle
\eea
is the two-quark component, and further terms in \eqref{mesonstate}
represent contributions from the multi-quark (i.e. the four-quark,
six-quark, etc) sectors. The two-quark approximation is developed
under the assumption that \eqref{psitwo} is the dominating component
of the meson state. It is certainly valid at sufficiently weak
coupling, when the parameter $\lambda$ is small,
but we will see that it provides meaningful description of
the spectrum even at very large values of this parameter. Anyway, the
two-quark approximation can be improved by taking the multiquark
components in \eqref{mesonstate} into account perturbatively (see
Sect.6). 

The two-quark component is completely characterized by the ``meson 
wave-function'' $\Psi(p_1,p_2)$ in \eqref{psitwo}. By
the definition \eqref{psitwo}\ the wave function is antisymmetric,
$\Psi(p_1,p_2) = -\Psi(p_2,p_1)$. If the multi-quark components in
\eqref{mesonstate} are neglected, the eigenvalue
problem $({\bf H}-E)\mid\Psi_P\,\rangle=0$ reduces to the integral
equation
\bea\label{inteqo}
\big[\varepsilon(p_1)+\varepsilon(p_2)-\Delta E\big]\Psi(p_1,p_2) =
&\nonumber\\
f_0\,\,
\dashint_{-\infty}^{\infty}\,2\pi\,\delta(p_1+p_2-q_1-q_2)&
{\cal G}(p_1,p_2|q_1,q_2)\,\Psi(q_1,q_2)\,{{dq_1}\over{2\pi}}
{{dq_1}\over{2\pi}}
\eea
where at this point $\varepsilon (p)$ stands for the free quark energy
$\varepsilon(p) = \omega(p)=\sqrt{m^2+p^2}$; we introduce this
new notation here in preparation to the discussion in Sect.4
below, where $\varepsilon(p)$ will become the ``dressed'' quark energy. In
\eqref{inteqo} $\Delta E = E - E_{\rm vac}$ is the energy above the ground
state, and 
\bea\label{fodef}
f_0 = 2{\bar\sigma}\,h = m^2\,\lambda
\eea
is the ``string tension''. In fact we will refer to the parameter
$f_0$, Eq.\eqref{fodef}, as the ``bare'' string tension, as opposed to
``dressed'', or ``effective'' string tension $f$ which will replace $f_0$
when radiative corrections (originating from the multi-quark components
in \eqref{mesonstate}) are taken into account, as will be discussed in
greater details in Sect.6. The kernel ${\cal G}$ in the
right-hand side is the matrix element \eqref{ffact}\,. The kernel is
singular, and the r.h.s. of \eqref{inteqo} involves the principal
value of the singular integral. For the meson state with the momentum
$P$ one takes the wave function of the form
\bea\label{fixedp}
\Psi(p_1,p_2) = (2\pi)\,\delta(p_1+p_2-P)\,\Psi_P (p_1-P/2)
\eea
with antisymmetric $\Psi_P(p) = -\Psi_P(-p)$. Then the equation
\eqref{inteqo} takes the form
\bea\label{inteq}
\big[\varepsilon\big(P/2-p\big)+\varepsilon\big(P/2+p\big)-\Delta
  E\big]
\Psi_P (p) =
f_0\,\,\dashint_{-\infty}^{\infty}\,G_P (\,p\,|\,q\,)\,\Psi_P
(q)\,{{dq}\over{2\pi}}\ ,
\eea
where the kernel $G_P (p|q)$ is the function \eqref{ffact} evaluated
at $p_1+p_2 = q_1+q_2 = P$,
\bea\label{gpgg}
G_P(p|q) \ =
\ {\cal G}\bigg({P\over 2}+p,\, {P\over 2}-p\,\bigg|\,{P\over 2}+q,\,{P\over
  2}-q\bigg)\ .
\eea
The meson wave-function $\Psi_P(p)$ is assumed to be normalizable
\footnote{More precisely, the states \eqref{psitwo} are assumed to be
  $\delta$-normalizable in terms of the meson momentum $P$. Since
$$
\langle\,\Psi_{P}\mid\Psi_{P\,'}\,\rangle =
2\pi\,\Delta E\,\delta(P-P\,')\,||\Psi_P||^2
$$
this implies the Eq.\eqref{twonorm}\,. The factor $\Delta E$ is
  inserted here in anticipation of relativistic normalization
of the meson states.},
\bea\label{twonorm}
||\Psi_P||^2 \ \equiv \  {1\over {2\,\Delta E}}
\,\int_{-\infty}^{\infty}\,|\Psi_P (p)|^2\,
{{dp}\over{2\pi}}\ \ < \ \infty\ ,
\eea
and the equation \eqref{inteq} is understood as the eigenvalue problem
for the meson energy $\Delta E$.

Observe that the kernel $G_P (p|q)$ in \eqref{inteq} has
the second-order poles at $q = \pm \,p$, with the residues $1$. More
precisely, one can check that
\bea\label{doublepoles}
G_P (p|q)\, = \,{1\over{(p-q)^2}}\, - \,{1\over{(p+q)^2}}\, +\, G_{P}^{({\rm
    reg})} (p|q)\ ,
\eea
where the last term is regular when both $p$ and $q$ take real
values. Therefore the principal value integral in the r.h.s. of
\eqref{inteq} has the same effect as the linear potential in \eqref{htwo},
provided the separation between the quarks is large enough, $m\,|{\rm
  x}_1 - {\rm x}_2| >> 1$. The short-range interaction described by the
regular term $G_{P}^{({\rm reg})}(p|q)$ is what makes \eqref{inteq}
different from the simple-minded Eq.\eqref{htwo}\ .

The integral equation \eqref{inteq} is the result of straightforward
application of the idea of the meson being the two-quark
construct. However, it is well known that in an interacting theory
neglecting the multi-quark components of the state \eqref{mesonstate}
generally violates Lorentz invariance. This is because in a
relativistic theory the quarks must be allowed to move occasionally
backward in time, the possibility depicted in Fig.2 but obviously
neglected when all the multi-quark terms in \eqref{mesonstate} are 
discarded. As the result, the equation \eqref{inteq} is not
Lorentz-invariant, in particular the eigenvalues $\Delta E$ are not
expected to have correct dependence $\sqrt{M^2+P^2}$ on the momentum
$P$.
\begin{figure}
\centering
\includegraphics[width=12cm]{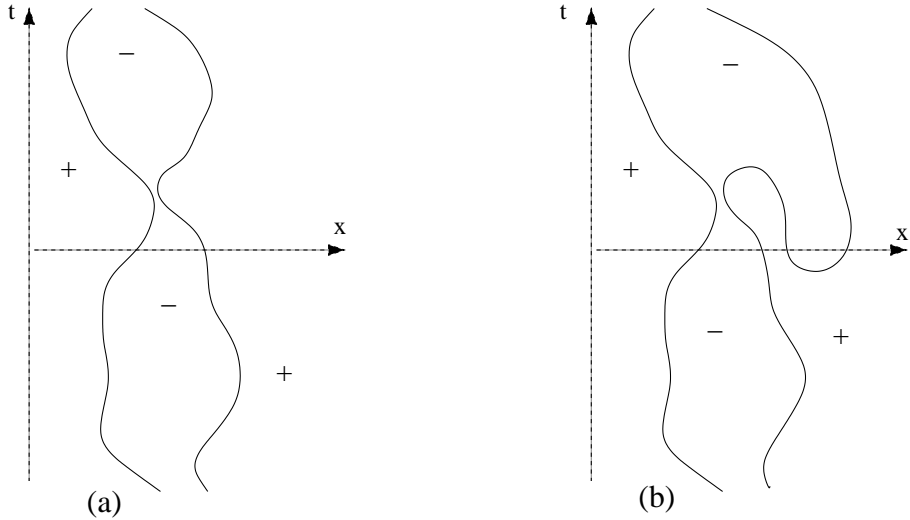}
\caption{\small Possible world lines of quarks in a meson. (a) Both quarks
propagate forward in time. (b) Creation an annihilation of virtual
pairs leads to the presence of more then two quarks in the
intermediate state.} 
\label{fig-roll}
\end{figure}

To restore the Lorentz invariance, it is necessary to retain certain
contributions from the multi-quark sectors in \eqref{mesonstate},
the ones associated with the backward-in-time propagations of the type
depicted in the Fig.2b. Adding such contributions would result in the
Lorentz invariant version of the two-quark approximation. Analogous
problem in perturbative quantum field theory is solved by the
Bethe-Salpeter equation (see e.g. \cite{qft}). Unfortunately, in the
IFT manifestly covariant perturbation theory in $h$ is not
available. For that reason, in the next Section we discuss the
multi-quark contributions associated with the
backward-in-time propagation of the quarks. These contributions are
described as corrections to the dispersion low of the quarks inside
the meson. We will argue that these corrections disappear in the limit
$P\to\infty$, and therefore the Lorentz-invariant Bethe-Salpeter
equation is essentially the equation \eqref{inteq} in the
infinite-momentum frame. Similar arguments were previously
put forward in the context of the 'tHooft model in \cite{bars}

\section {The Bethe-Salpeter equation}\label{BS}

As usual, the integral equation \eqref{inteq} can be understood as a
device for summing up certain class of diagrams of the perturbation
theory in $h$. Here and below we have in mind usual quantum-mechanical
perturbation theory (see e.g. \cite{landau}), not the covariant
Feynman perturbation theory \footnote{As was already said, manifestly
covariant perturbation theory in $h$ is not yet developed. See however 
calculations in Ref.\cite{mccoy3} which seem to be suggestive in this
respect.}, so that all the intermediate lines in 
the diagrams below are on-shell. The equation \eqref{inteq} takes into
account the ``ladder'' diagrams shown in Fig.3a, where the blobs
represent the matrix elements \eqref{matrix}, with only two
quarks involved in the intermediate states. But the Eq.\eqref{inteq}
neglects all the diagrams with four, six, or more intermediate quarks,
like the one shown in the Fig.3b.
 \begin{figure}
\centering
\includegraphics[width=12cm]{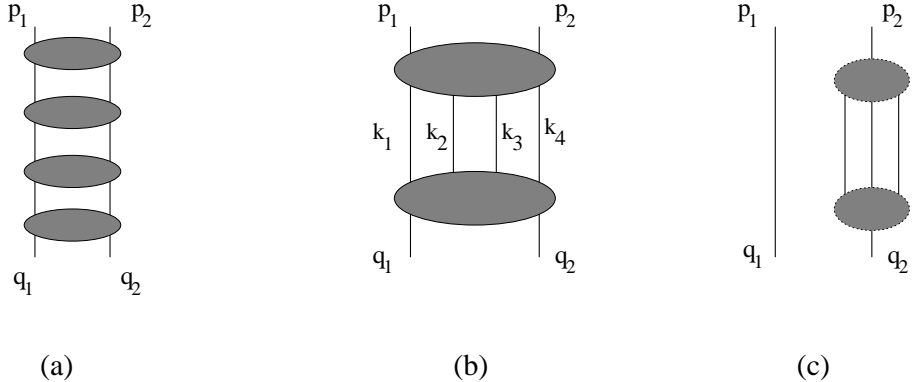}
\caption{\small Diagrams of the perturbation theory in $h$. (a) The
  ``ladder'' diagrams. The blobs represent the matrix elements
\eqref{matrix}\ . These diagrams are summed up by the
  Eq.\eqref{inteq}. (b) The diagram with four intermediate quarks.
(c) Example of a disconnected part of the diagram in (b).}
\label{twoquark}
\end{figure}
The blobs in the Fig.3b stand for
the matrix elements between the two-quark and four-quark states, and
the integration over the intermediate momenta $k_1, k_2, k_3, k_4$
(subject to the constraint $k_1+k_2+k_3+k_4=p_1+p_2 = q_1+q_2$) is
implied. The contribution of the diagram in Fig.3b contains 
disconnected parts, in which only one of the particles is affected by
the interaction, as is depicted by the diagram in Fig.3c. In the
disconnected parts like this the momentum conserves separately within
each disconnected component, i.e. such diagrams contain two momentum
\begin{figure}
\centering
\includegraphics[width=12cm]{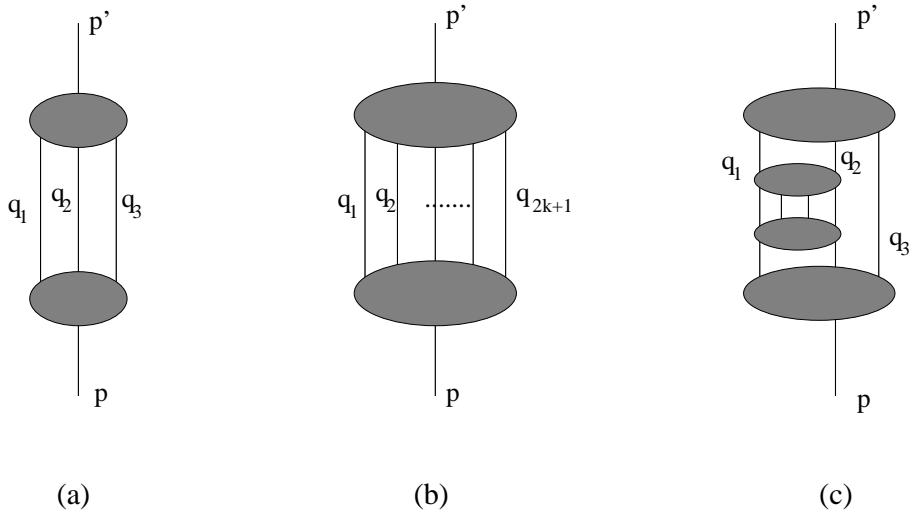}
\caption{\small The quark self-energy: (a) The simplest self-energy diagram
of the order $\sim h^2$; its contribution $\Sigma_{2}^{(3)}(p)$
appears in the Eq.\eqref{disconnected}. (b), (c) Examples of more
complicated self-energy diagrams.}
\label{selfenergy}
\end{figure}
$\delta$-functions. In particular, the contribution of the diagram
in Fig.3c to the energy can be written as
\bea\label{disconnected}
\bigg[{\rm {{disconnected\ \ }\atop{diagram\ in \ Fig.3c}}}\bigg] \ =
(2\pi)^2\,\delta(p_1 -q_1)\delta(p_2-q_2)\,\,\Sigma^{(3)}_2 (\,p_2\,)\ ,
\eea
where $\Sigma_{2}^{(3)} (p)$ is the self-energy term from the diagram in
Fig.4a. Clearly, \eqref{disconnected} and similar disconnected
contributions involving the self-energy diagrams in Fig.4 can be
accounted for by replacing the factors $\omega(p) = \sqrt{m^2+p^2}$
in the left-hand side of the Eq.\eqref{inteqo} by the dressed quark
energy, $\omega(p)\ \to \ \varepsilon(p)\ = \
\omega(p)\,+\,\Sigma(p)$, where the self-energy correction $\Sigma(p)$
in principle can be determined
order by order in the perturbation theory in $h$. On general grounds,
one expects to have two kinds of perturbative contributions to the
self-energy. As usual, there are contributions associated with the
one-particle irreducible diagrams. These corrections do not modify the
relativistic form \eqref{omegadef} of the momentum dependence of
the quark energy, while making corrections to the quark mass $m$. At
this stage we shall ignore the radiative corrections to the quark 
mass\,\footnote{The leading correction $\sim h^2$ to the quark was
determined in Ref.\cite{fz2} (the result is quoted in
Eqs.\eqref{mqseries}, \eqref{mqtwo} below). It may be useful to note that the
calculations involve subtracting the one-particle
reducible parts (Eqs. (5.10),(5.13), and Fig.3 of Ref.\cite{fz2}), which are
exactly the diagrams Fig.5a and Fig.5b here. The leading correction
in \eqref{epsilonc} below comes from adding the diagram in Fig.5b back.},
postponing the discussion to Section 6. Here we are more
interested in the contributions from the one-particle reducible
diagrams. Simplest example of the one-particle reducible
diagram is shown in Fig.5b, which can be interpreted as the quark
traveling backward in time in between two interaction events. Note
that unlike the one-particle reducible diagram with the forward time
propagation in Fig.5a, which is already accounted for in the sum of
the ladder diagrams in Fig.3a, the diagram in Fig.5b has to be taken
into account as separate contribution to $\varepsilon(p)$.
The difference between the diagrams
\begin{figure}
\centering
\includegraphics[width=12cm]{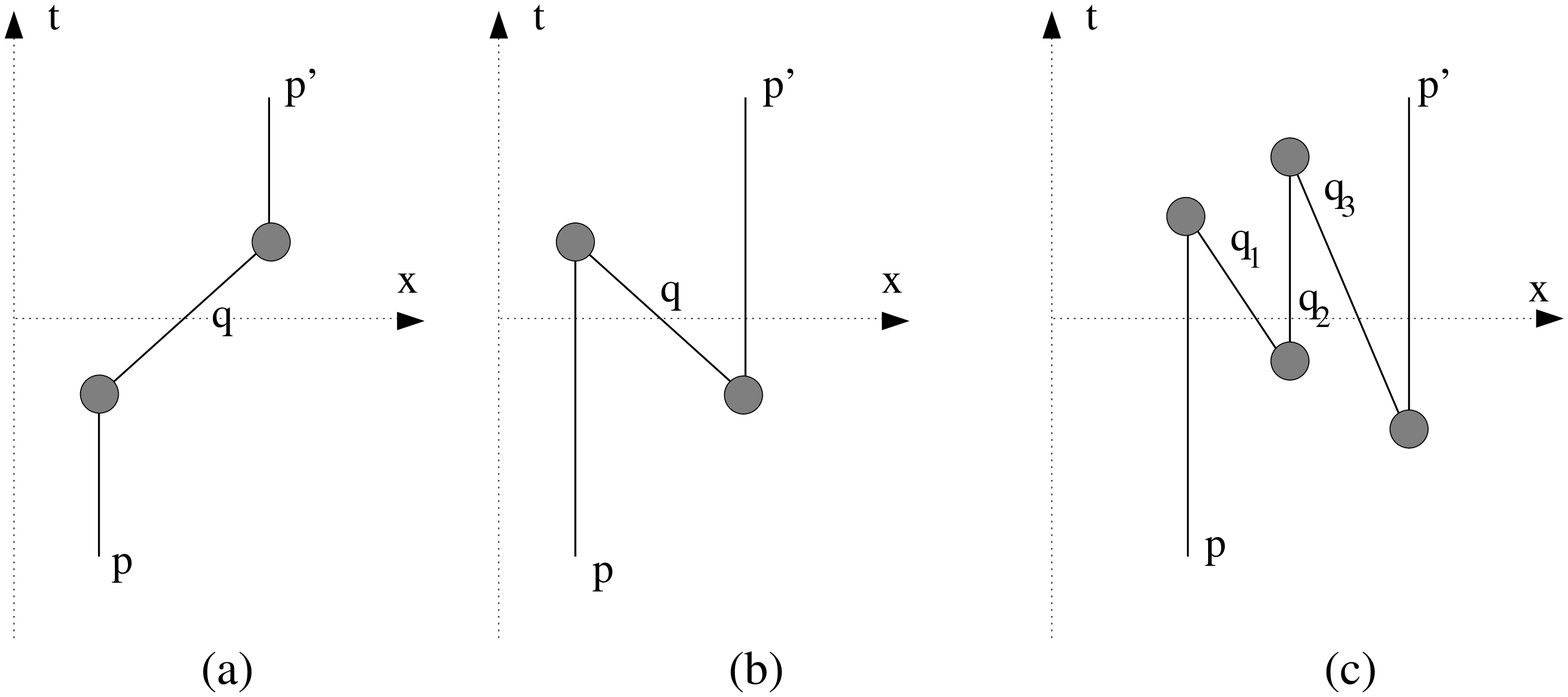}
\caption{\small One-particle reducible self-energy diagrams. Parts (a) and
(b) show two types of the order $h^2$ diagrams, with one and three
quarks in the intermediate state, respectively. Part (c) shows
an example of the higher-order diagrams which contribute to
$\varepsilon(p)$.}
\label{vareps}
\end{figure}
Fig.5a and Fig.5b is in the time order of the interaction events;
since this distinction may depend on the Lorentz frame, the contribution
of the diagram in Fig.5b is not expected to be Lorentz covariant. Similar
non-covariant corrections to the quark self-energy arise from the
one-particle reducible diagrams of higher orders with the
backward-in-time propagation (e.g. the diagram in Fig.5c). The quark
energy $\varepsilon(p)$ includes all the one-particle reducible
diagrams which are not reproduced by iterations of the equation
\eqref{inteq} (such are the diagrams in Figs.5b and 5c, but not in
Fig.5a).

It is instructive to calculate explicitly the lowest-order correction
term in $\varepsilon(p)$, This term comes from the diagram in Fig.5b,
which reads
\bea\label{deloma}
-\,h^2\,(2\pi)\delta(p-p')\,
\int_{-\infty}^{\infty}\,{{(2\pi)\delta(p+q)\,
|\langle\,0\mid \sigma(0)\mid
p,q\,\rangle|^2}\over{\omega(p)+\omega(q)}}\,{{dq}\over{2\pi}}\ .
\eea
Performing the trivial integration, and using explicit form
\cite{karowski} of the matrix element
\bea\label{ftwo}
\langle\,0\mid \,\sigma(0)\,\mid p,\,q\,\rangle =
{{i\,{\bar\sigma}}\over{\sqrt{\omega(p)\,\omega(q)}}}\,\,
{{p-q}\over{\omega(p)+\omega(q)}}
\eea
one finds
\bea\label{epsilonc}
\varepsilon(p)\ =\ \omega(p)\,-
\,{{\lambda^2}\over 8}\,{{p^2}\over{\omega^5 (p)}} + O(\lambda^4)\ ,
\eea
where $\lambda$ is defined in Eq.\eqref{lambdadef}. It is possible to
show that this correction in \eqref{epsilonc} is exactly what is
needed in order to restore the Lorentz covariance of the spectrum of
the equation \eqref{inteqo}\,, to this order in $\lambda$. The
calculation supporting this statement is rather involved; we will
present it elsewhere.

In principle, the higher order terms in \eqref{epsilonc} can be
determined by direct calculation of the corresponding diagrams.
It is reasonable to expect that incorporating these terms in
$\varepsilon(p)$ will restore the Lorentz covariance of the equation
\eqref{inteqo} order by order in $\lambda^2$ (analogous statement for the
't Hooft's model has been proven in Ref.\cite{bars}). In particular, with all
contributions to $\Sigma(p)$ taken into account, the $P$-dependence of all the
eigenvalues $\Delta E$ of \eqref{inteqo} must assume the relativistic form
\bea\label{deltae}
\Delta E = \sqrt{M^2 + P^2}
\eea
with $M$ interpreted as the meson mass. Let us stress that at the
moment we have no proof of this statement beyond the order
$\lambda^2$. We intend to come back to this question in the future.
In the remaining part of this paper we proceed under this assumption. 

Once the assumption is accepted, there is an obvious way to circumvent 
calculation of $\Sigma(p)$ altogether. Note that the leading term $\sim
\lambda^2$ in \eqref{epsilonc} decays fast (as $|p|^{-3}$) when
$p\to\infty$. It is possible to verify that the higher-order
terms in \eqref{epsilonc} decay yet faster, so that the asymptotic
form
\bea\label{epsas}
\varepsilon(p) = |\,p\,| + {{\ m^2}\over{2 \,|\,p \,|}} +
O\bigg(\,{1\over{|\,p\,|^3}}\,\bigg)
\eea
is valid to all orders in $\lambda\,$\,\footnote{Here we ignore all 
corrections to the quark mass determined by the one-particle
irreducible self energy diagrams. Of course, exact asymptotic form 
of $\varepsilon(p)$ is given by the Eq.\eqref{epsas}\ ,  with $m$ replaced 
by the ``dressed'' quark mass $m_q$.}. Consider the 
eigenvalue equation \eqref{inteq} and take the limit $P\to +\infty$. 
As is shown in Appendix A, in this limit, and under appropriate 
normalization, the meson
wave function $\Psi_P (p)$ remains finite at $2|p|/P < 1$, but vanishes
as $1/P^2$ outside this domain. For $2|p|/P < 1$ both
sides of the equation \eqref{inteq} decay as $1/P$ at
large $P$. Note that this leading term of the $P\to \infty$ asymptotic
is not affected at all by the correction terms in \eqref{epsilonc},
since $\Sigma(p)$ decays faster then $1/p$. Balancing the
coefficients in front of $1/P$ asymptotic in both sides of
\eqref{inteq} yields a nontrivial eigenvalue equation for the
parameter $M^2$ in \eqref{deltae}\,. This equation is derived
in Appendix A. Its most convenient form is obtained by changing
to the rapidity variables
\bea\label{raps}
2p = P\,\tanh\theta\ , \qquad 2q = P\,\tanh\theta'
\eea
before sending $P$ to infinity, and introducing the notation
$\Psi(\theta)$ for the limit of $\Psi_P (p)$ at $P\to\infty$. It is
\bea\label{bs}
\bigg[{m^2} -
{{M^2}\over{4\,\cosh^2\theta}}\bigg]\Psi(\theta) =
f_0\,\,\dashint_{-\infty}^{\infty}\,G(\theta|\theta')\,
\Psi(\theta')\,
{{d\theta'}\over{2\pi}}\ ,
\eea
where
\bea\label{kernelt}
G(\theta|\theta') =
2\,{{\cosh(\theta-\theta')}\over{\sinh^2(\theta-\theta')}} +
  {1\over 4}\,{{\sinh\theta}\over{\cosh^2\theta}}\,
{{\sinh\theta'}\over{\cosh^2\theta'}}\ .
\eea
We will refer to \eqref{bs} as the
Bethe-Salpeter (BS) equation.

Let us make few general remarks on the properties of the BS equation
\eqref{bs}. First, the wave function $\Psi(\theta)$ should be
regarded as a vector in the Hilbert space with the metric
\bea\label{metric}
||\Psi||^2 =
\int_{-\infty}^{\infty}\,{{|\Psi(\theta)|^2}\over{4\,\cosh^2\theta}}
\,{{d\theta}\over{2\pi}}
\eea
which is just the metric \eqref{twonorm} rewritten in terms of the
rapidity variable $\theta$, in the limit $P=\infty$. Second,
the equation is understood as the eigenvalue problem for
the parameter $M^2$,
\bea\label{bsh}
{\hat H}\Psi = M^2\Psi\ ,
\eea
where the operator ${\hat H}$, defined as
\bea\label{hhat}
{\hat H}\Psi (\theta) = 4\,\cosh^2\theta\,\bigg[m^2\,\,
\Psi (\theta) - f_0\,\,
\dashint_{-\infty}^{\infty}\,G(\theta|\theta')\,\Psi(\theta')\,
{{d\theta}\over{2\pi}}\bigg]\ ,
\eea
is Hermitian with respect to the metric \eqref{metric}\ . Its
eigenvalues are real and positive. We will use the notations 
${\widetilde{M}}_{n}^2$ with $n =1,2,3,\ldots$ for the
successive eigenvalues of \eqref{bsh} (arranged in the order of 
increasing magnitude) and $\Psi_n(\theta)$ for the corresponding 
eigenfunctions. The quantities ${\widetilde M}_n$ provide certain
approximations for the actual meson masses, and we reserve the
notation $M_n$ for those actual masses.  

\section{Weak coupling expansions}

If the magnetic field $h$ is small, such that $f_0 << m^2$, the 
Bethe-Salpeter equation \eqref{bs} reproduces physics of
weakly confined quarks, as described in Sect.2. According to this
picture, one expects that in the limit $\lambda\to 0$ each individual 
eigenvalue ${\widetilde M}_{n}^2$ approaches $4 m^2$ from the above, so 
that the full spectrum $\{{\widetilde M}_{n}^2 \}$ of the equation
\eqref{bsh} becomes dense in the segment $[4 m^2, \infty)$. Here we 
develop the weak coupling expansions of 
the eigenvalues ${\widetilde M}_{n}^2$. The nature of the expansion
depends on what part of the spectrum $\{{\widetilde M}_{n}^2\}$ one is 
interested in. First, one might be interested in the small-$\lambda$
expansion of an individual eigenvalue ${\widetilde M}_{n}^2$, with fixed $n$.
Since the kinetic and potential energies of the quarks in any given
meson state become small at small $\lambda$, we refer to this case
as the {\it low energy expansion}. The equation \eqref{mccoys} is just
the first therm of this expansion, and in general the low-energy
expansion of ${\widetilde M}_{n}^2$ turns to be in fractional powers of $\lambda$,
namely in the powers of the parameter
\bea\label{tdef}
t = \lambda^{1\over 3}\ .
\eea
Second, the behavior of the higher levels ${\widetilde M}_{n}^2$ with 
$n \gtrsim {1\over \lambda}$ may be of interest; in this case one can 
derive the {\it semiclassical expansion} extending the
Eq.\eqref{wkbmass}\,. This expansion is in integer powers of
$\lambda$, and the Bohr-Sommerfeld equation \eqref{wkbmass} appears as 
its leading term, while the higher-order corrections were discussed in
\cite{rut}. In all cases it is useful to remember
that the BS equation itself is an approximation, and therefore
starting from certain order ($t^9$ in the low-energy and
correspondingly $\lambda^3$ in the semiclassical expansion) both
expansions exceed the accuracy of the BS equation \eqref{bs}. From
these orders on, the coefficients of the weak coupling expansions of 
the actual masses $M_n$ start to receive contributions associated with the 
multi-quark components of the meson states\,\footnote{Here we
refers to the multi-quark corrections which cannot be absorbed into
the renormalizations of the parameters $m$ and $f_0$. Corrections to
these parameters appear at lower orders; more details are in Section
6.}\ . In this Section we ignore such corrections, but we will come
back to them in Sect.6. 

We develop the weak coupling expansions starting from
the following approximation for the solution of the equation \eqref{bsh}\ ,
\bea\label{psioo}
\Psi^{(0)} (\theta) =
\sinh\theta\,\,\int_{-\infty}^{\infty}\,
{{e^{{i\over\lambda}\,S(\beta)}\,\,\,\cosh\beta\,\,d\beta}\over
{\sinh(\theta+\beta-i0)\,\sinh(\theta-\beta+i0)}}\ , 
\eea 
where
\bea\label{soaction} 
S(\beta) = {{M^2}\over{4
m^2}}\,\tanh\beta\,-\,\beta\ .
\eea
It is easy to show that
$\Psi^{(0)}(\theta)$ is normalizable (with respect to the metric
\eqref{metric}) at any value of $M^2$ . The
quality of this approximation is determined by the function 
$\Delta(\theta | M^2)$ defined as (in what follows we usually suppress 
the argument $M^2$)
\bea\label{deltadef}
4 m^2\,\sinh\theta\ \Delta(\theta)\ \equiv\ \big[{\hat H} -
M^2\big]\,\Psi_0 (\theta)\ ,
\eea
where ${\hat H}$ is the operator \eqref{bsh}. By
straightforward transformations (which we sketch in the Appendix B)
this function can be written as the integral
\bea\label{deltaint}
{\Delta}(\theta) =
\int_{-\infty}^{\infty}\,\bigg[{{M^2}\over{4 m^2\,\cosh\beta}}+
  {{i\lambda}\over
    8}\,{{\sinh\beta}\over{\cosh^2\beta}}+
\lambda\,\,B(\theta |\beta)\bigg]\,
e^{{i\over\lambda}\,S(\beta)}\,d\beta\ ,
\eea
where
\bea\label{beven} B(\theta|\beta) = {1\over\pi}\,
{d\over{d\beta}}\bigg[{{\cosh^2\theta}\over{\sinh\theta}}\,
\bigg({{(\theta+\beta)}\over{\sinh(\theta+\beta)}} -
{{(\theta-\beta)}\over{\sinh(\theta-\beta)}}\bigg) - {1\over
  4}\,{{\beta}\over{\cosh\beta}}\bigg]\ .
\eea
Note that the terms $\sim\lambda$ in the integrand in \eqref{deltaint}
are sorted out according to their symmetry with respect to
$\beta\to-\beta$. 

Next, we develop a small\,-\,$\lambda$ expansion of the integral
\eqref{deltaint}. Let us note that the integrand in
\eqref{deltaint}\ is independent of $\theta$ except for the last
term $B(\theta|\beta)$ in the brackets there. If the $\theta$
dependence could be ignored, it would be possible to turn the whole 
integral \eqref{deltaint} to zero by tuning the parameter $M^2$ to 
special values, the zeros of the function $\Delta(M^2) \equiv 
\Delta(\theta | M^2)$; these special values would then determine the 
eigenvalues ${\widetilde M}_{n}^2$. It turns out that the term
$B(\theta|\beta)$ in \eqref{deltaint} does not contribute to the 
expansion all the way up to rather high order - up to $t^{8}$ of 
the low-energy expansion and up to $\lambda^2$ in the semiclassical 
expansion - and therefore evaluating first few terms of the weak 
coupling expansion from \eqref{deltaint} \ is very easy. At higher 
orders the $\theta$-dependence shows up and hence no choice
of $M^2$ can turn the function \eqref{deltaint} to identical zero. 
This is well expected - after all \eqref{psioo} is an approximate, not
exact solution of the Eq.\eqref{bs}. Therefore finding the
terms of the order $\lambda^3$ and higher requires additional 
ingredients. Let us show how to develop systematic weak coupling 
expansion of the eigenvalues ${\widetilde M}_{n}^2$ to any 
order\,\footnote{In the case under consideration the higher order terms have
somewhat limited value: as we explain in Section 6 such terms
compete with the multi-quark corrections which are generally beyond
control of the BS equation \eqref{bs}. Here we still do it as a
technical exercise, since the method can be applied in other models,
notably in the 't Hooft's model of multicolor QCD \cite{thooft}, where
very similar Bethe-Salpeter equation (shown as the Eq's
\eqref{bsu},\eqref{thooftker} in Appendix A) is exact.}. 

We start with the trivial observation that any
eigenvalue ${\widetilde M}_{n}^2$ satisfies the equation
\bea\label{cneq}
C_n ({\widetilde M}^2) = 0\ ,
\eea
where
\bea\label{cdef}
C_n (M^2) = \big(\Psi_n , \big[{\hat
H}-M^2\big]\Psi^{(0)}\big)\ .
\eea
Here and below the brackets
$(\ ,\ )$ denote the scalar product associated with the metric
\eqref{metric} . We also assume the eigenvectors to be
orthonormalized, $(\Psi_n, \Psi_k) = \delta_{n,k}$. Note that
\bea\label{cnint}
C_n(M^2) = \int\,{{
    m^2\,\sinh\theta\,\,\Delta(\theta)\,
\Psi_n(\theta)}\over{\cosh^2 \theta}}\,{{d\theta}\over{2\pi}}\ .
\eea
At small
$\lambda$ the function $\Psi^{(0)} (\theta)$ provides good approximation
for $\Psi_n (\theta)$, and moreover one can find corrections by
iterating the equations
\bea\label{psipert}
\Psi_n (\theta) = {1\over{\big(\Psi^{(0)},\Psi_n\big)}}\,\,
\bigg[\Psi^{(0)}(\theta) -
\sum_{k\neq n}\,{{C_k (M^2)}\over{M_{k}^2 -
      M^2}}\,\Psi_k (\theta) \bigg]_{M^2 = M_{n}^2}\ ,
\eea
Then the solution has to be plugged back into \eqref{cdef}, providing
further corrections to the eigenvalues through the Eq.\eqref{cneq}.

We illustrate this technique by explicit calculations of the
first few orders, both in the low-energy and semiclassical expansions.

\subsection{Low energy expansion}

Assume that the parameter $M^2$ in \eqref{deltaint} is close to $4
m^2$, so that $M^2 - 4\,m^2\ \lesssim \ t^2$. Note that any given
eigenvalue ${\widetilde M}_{n}^2$ enters this domain at sufficiently
small $\lambda$. When this is the case, the integral \eqref{deltaint}
is dominated by contribution from $\beta \sim t$. It is convenient
then to change to a new variable $u = -\beta/t$\ . We also write
${{M^2}} = 4 m^2\,(1 + z\,t^2 + O(t^4))$ with some coefficient $z$
to be determined below. In the leading order in $t$ the
Eq.\eqref{deltaint} reduces to the integral defining the Airy function,
\bea\label{airy}
\Delta(\theta)\ =\ t\,\,\int_{-\infty}^{\infty}\, e^{{i\over
3}{u^3} - iz u} \,du + O(t^3)\ .
\eea
Hence, it is sufficient to set the parameter $-z$ equal to any zero
of the Airy function,
\bea\label{zdef}
{\rm Ai}(-z) =0\,,
\eea
to make \eqref{airy} vanish to the leading order in $t$. This
reproduces the equation \eqref{mccoys}. In what follows we use the
notation $z$ for a generic solution of the equation \eqref{zdef}\,.

More generally, we look for the solutions of \eqref{cneq} in the form
\bea\label{lowexp}
{{{\widetilde M}^2}\over{4 m^2}} =
1 + z\,t^2 + \sum_{k=3}^{\infty}\,\epsilon_k\,t^k\ .
\eea
The integral \eqref{deltaint} can be evaluated order by order in $t$. 
Let us write its expansion as
\bea\label{deltaser}
{{\Delta(\theta)}\over{\pi\,{\rm Ai}' (-z)}}\  = \ D_3
(\theta)\,t^3 + D_4 (\theta)\,t^4 + D_5 (\theta)\,t^5 + \cdots\,.
\eea
A straightforward calculation\footnote{The calculation involves the
integrals 
\bea\label{ik}
\int_{-\infty}^{\infty}\,(i u)^k\,e^{{i\over 3}{u^3} - iz u} \,du \
\equiv \ \pi\,{\rm Ai}' (-z)\ I_{k}.  \nonumber
\eea
The factors $I_k$ in the r.h.s. of this equation are easily
computed one by one using the recursion
\bea\label{recurrent} I_{k+2} = -z\,I_k + k\,I_{k-1} \qquad {\rm with}
\qquad I_0 = 0 \quad {\rm and}\quad I_1 =1\ , \nonumber
\eea
where the last two equations follow from
Eqs.\eqref{zdef}\,\eqref{ik}\,.} yields
\bea\label{dk}
D_3(\theta) = -\,\epsilon_3\ , \qquad D_4(\theta) =
{{z^2}\over 5}\,-\,\epsilon_4\ , \qquad D_5 (\theta) = - \epsilon_5\ ,
\qquad {\rm etc}
\eea
It turns out that the coefficients $D_k (\theta)$ have no
$\theta\,$-\,dependence all the way up to $D_{8}(\theta)$, so that they can be
turned to zero by suitable choice of $\epsilon_3,\, \epsilon_4,\,
\cdots, \, \epsilon_{8}\,$, e.g. $\epsilon_3 =0$, $\epsilon_4 = z^2/5$, 
etc. It also turns out that the coefficients $\epsilon_5$ and $\epsilon_7$
obtained in this manner vanish (but higher order odd terms are present in
\eqref{lowexp}, see Eq.\eqref{lowseriesa} below). Thus we find
\bea\label{lowseries} {{{\widetilde M}^2}\over{4 m^2}} = 1 + z \,t^2 +
{{z^2}\over{5}}\,t^4 - \bigg({{3\,z^3}\over 175}+{57\over
280}\bigg)\,t^6 + \bigg({{23\,z^4}\over 7875} +
{{1543\,z}\over{12600}}\bigg)\,t^8 + \ldots
\eea
The first four terms here are identical to the results previously
reported in Ref.\cite{fz1}\,\footnote{Our calculations in
Ref.\cite{fz1} were performed using the Eq.\eqref{inteq}, with
$\varepsilon(p) = \omega(p)$, in the center-of-mass frame $P=0$. 
Corrections to $\varepsilon(p)$ contribute to ${\widetilde M}_{n}^2$
starting only from the order $t^8$; that is why all terms quoted
in \cite {fz1} are correct.}, but the fifth ($\sim t^8$) term is new.
Starting with the term $\sim t^9$ the $\theta$-dependent term 
$B(\theta|\beta)$ in \eqref{deltaint} becomes relevant, and the 
coefficients $D_k(\theta)$ with $k\geq 9$ generally
have nontrivial dependence on $\theta$. For example
\bea\label{dnine}
D_{9}(\theta) = {3\over 8}\,B_0 (\theta) + {4\over 3}\,B_2 (\theta) +
2\,B_4 (\theta) - \epsilon_9\ ,
\eea
where the functions $B_{2n}(\theta)$ are the coefficients in the
expansion
\bea\label{bexpan}
B(\theta|\beta) = B_0 (\theta) + B_{2}(\theta)\,\beta^2 +
B_{4}(\theta)\,\beta^4 + \ldots
\eea
At this order and beyond we have turn to the equation \eqref{cneq}, 
and determine the coefficients $\epsilon_9,\ \epsilon_{10}, \ $ etc 
from the condition that the integral \eqref{cnint} vanishes order 
by order in $t$. We do not enter the details of this calculation here, 
only remark that the integral \eqref{cnint} receives the main
contribution from the domain $\theta \sim t$, therefore its expansion 
in $t$ involves the expansions of the coefficients $D_9 (\theta), \ 
D_{10}(\theta), \ \ldots$ in the powers of $\theta$\,. Let us quote
few more terms of the expansion \eqref{lowexp},
\bea\label{lowseriesa}
\epsilon_9 = {{13}\over{1120\, \pi}}\ , \quad \epsilon_{10} =
-{{1894\,z^5}\over{3031875}} - {{23983\,z^2}\over{242550}}\ , \quad
\epsilon_{11} = {{3313\,z}\over{10080\,\pi}}\ .
\eea

Quick inspection of \eqref{lowseries} and \eqref{lowseriesa} (as well
as the higher order terms) reveals a general structure. The
coefficients $\epsilon_k$ are polynomials in $z$, of growing
degree. The even coefficients have the form $\epsilon_{2l}=  a_l\,z^l
+ b_l\,z^{l-3} +\cdots$, while for the odd terms we have
$\epsilon_{2l+1} \sim z^{l-4}$. The parameter $z$ becomes large for
higher eigenvalues (the Airy zeros grow as $z_n \sim n^{2\over 3}$).
If $t << 1$ but $z_n\,t^2 \sim 1$ or greater, it is useful to sum up 
the leading terms $\sim \big(n^{1\over 3}\,t\big)^{2l}$. It is not
hard to see that this is exactly what the Bohr-Sommerfeld equation 
\eqref{bscondition}, \eqref{wkbmass} accomplishes. Then one can try
to collect the subleading terms $\sim t^6\,\big(n^{1\over
  3}\,t\big)^{2l}\,$, $\sim t^9\,\big(n^{1\over 3}\,t\big)^{2l}\,$,
etc. This is done by evaluating the higher order terms of the 
semiclassical expansion, as we discuss next.

\subsection{Semiclassical expansion}

If $\lambda <<1$ and $M^2$ is not too close to $4m^2$ the integral 
\eqref{deltaint} can be evaluated by the stationary phase method. 
There are two saddle points at $\beta = \pm\,\vartheta$, where
$\vartheta$ is a positive parameter defined by the equation
\bea\label{varthetadef}
M^2 = 4 m^2 \,\cosh^2 \vartheta\ .
\eea
Correspondingly, the integral can be written as
\bea\label{deltapm}
\Delta(\theta) = \Delta_{+}(\theta) + \Delta_{-}(\theta)
\eea
where the two terms represents the
contribution of the two saddle point $\beta=\pm\,\vartheta$,
respectively. Obviously, $\Delta_{-}(\theta) =
\Delta_{+}^{*}(\theta)$, so we can concentrate attention on one of
this terms, say $\Delta_{+}(\theta)$. In the leading approximation
only the first term in the square brackets \eqref{deltaint}
contributes, and it has to be evaluated at the saddle point $\beta =
\vartheta$. Thus in the leading order $\Delta(\theta)$ is independent
of $\theta$, namely
\bea\label{wkba}
{{\Delta^{({\rm classical})}(\theta)}\over{\sqrt{2\pi\lambda\sinh 
2\vartheta}}} =
\cos\bigg({{{\bar S}(\vartheta)}\over\lambda} - {\pi\over 4}\bigg)\ .
\eea
where
\bea\label{wkbaction}
{\bar S}(\vartheta) \equiv S(\beta)\big|_{\beta=\vartheta} = 
{{\sinh 2\vartheta - 2\vartheta}\over 2}
\eea
(${\bar S}(\vartheta)$ is not the same function as $S(\vartheta)$ 
because $\vartheta$ enters also through \eqref{varthetadef}). 
Vanishing of \eqref{wkba} leads exactly to the quantization condition
\eqref{bscondition}\,.

Corrections to the leading WKB formula can be obtained by developing
the standard loop expansions of the integral \eqref{deltaint} around
the saddle points. In the one-loop order the $\theta$-dependent term
$B(\theta|\beta)$ still is not too important (its effect is in
bringing in an overall factor $1 + \lambda\,B(\theta|\vartheta)$ in
$\Delta(\theta)$), and $\Delta(\theta)$ still can be turned to
identical zero by appropriate adjustment of $\vartheta$. Explicit
calculation yields
\bea\label{wkbc}
{{\Delta^{({\rm
    one-loop})}_{+}(\theta)}\over{\sqrt{2\pi\lambda\sinh
      2\vartheta}}}\,\,{2\over{1 + \lambda\,B(\theta|\vartheta)}} =
e^{-{{i\pi}\over 4}}\,e^{{i\over\lambda}{\bar S}(\vartheta)} \,
\bigg(1 + i\lambda\,{\bar S}_1 (\vartheta) + O(\lambda^2)\bigg)
\eea
where
\bea\label{wkbd}
{\bar S}_1 (\vartheta) =
-{1\over{\sinh 2\vartheta}}\,\bigg[{5\over
    24}\,{1\over{\sinh^2\vartheta}} + {1\over 4}\,{1\over{\cosh^2\vartheta}}
  - {1\over 12} - {1\over 6}\,\sinh^2\vartheta\bigg]
\eea
This leads to the one-loop correction to the quantization condition
\eqref{bscondition},
\bea\label{bsccc}
{{\sinh 2\vartheta_n - 2\vartheta_n}} =
2\pi\,\lambda\,\big(n-1/4) - \lambda^2\,{\bar S}_1 (\vartheta_n) - 
O(\lambda^3)\ .
\eea
The eigenvalues ${\widetilde M}_{n}^2$ are still related to $\vartheta_n$ 
through \eqref{varthetadef}, i.e. 
\bea\label{wkbe}
{\widetilde M}_{n}^2 = 4 m^2\,\cosh \vartheta_n
\eea

Let us remark here on the relation to the results of
Ref.\cite{rut}, where the semiclassical expansion to the same order
$\sim \lambda^2$ is developed on the basis of the two-quark
approximation in the center of mass frame. Our result \eqref{bsccc} is
similar but different from that of \cite{rut}. The calculation in
\cite{rut} ignores the correction $\sim\lambda^2$ to the quark self
energy, which is responsible for the Lorentz covariance of our
approach. It is easy to check that when the self-energy correction
\eqref{epsilonc} is added, the result of \cite{rut} becomes identical to
\eqref{bsccc}, \eqref{wkbe} above.

The higher-loop corrections generally leave the Eq.\eqref{wkbe}
unchanged while adding the higher-order terms
\bea\label{bsccd}
- \lambda^3\,{\bar S}_2 (\vartheta_n) - \lambda^4 \,
{\bar S}_3 (\vartheta_n) - \ldots\ ,
\eea
to the r.h.s. of the Eq.\eqref{bsccc}. Calculation of the coefficients 
${\bar S}_2 (\vartheta), {\bar S}_3(\vartheta), \cdots$ requires (rather
straightforward) evaluation of the
higher-loop contributions to $\Delta_{+}(\theta)$ and
$\Delta_{-}(\theta)$ in \eqref{deltapm}. At two loops (and of course
at all higher orders as well) $\Delta(\theta)$ acquires an essential 
dependence on $\theta$, which makes it impossible to turn it to
identical zero by an adjustment of $\vartheta$. Just like in our
previous analysis of the low-energy expansion, one finds the
higher-loop terms in \eqref{bsccd} from the equation \eqref{cneq}. 
Since $\Delta(\theta)$ remains slow-varying function at small
$\lambda$, the integral in \eqref{cnint} is
dominated by the stationary-phase points $\theta =
\pm\,\vartheta$ of the function $\Psi_n (\theta)$, and can be
evaluated order by order in the associated loop expansion. Thus, at
the two-loop level in the mass spectrum calculation, the leading-order
stationary-point calculation of \eqref{cnint} is appropriate,
therefore the coefficient ${\bar S}_2 (\vartheta)$ in \eqref{bsccd} is found
from the equation
\bea\label{twoloopeq}
\Delta^{({\rm two-loop})} (\theta)\big|_{\theta=\vartheta} =0\ .
\eea
Calculations of yet higher coefficients $S_3 (\vartheta), \ S_4
(\vartheta),$ etc involves also iterations of \eqref{psipert}, along 
with the higher-loop evaluation of the integral \eqref{cnint}. We did 
not perform these straightforward calculations explicitly because, as
was mentioned above, all terms beyond $\lambda^2$ in \eqref{bsccc}
exceed the accuracy of the BS equation itself - starting from the term
$\lambda^3$ the expansion of the actual masses $M_{n}$ receives
nontrivial contributions from the multi-quark sectors, and without
good control of these corrections the semiclassical analysis at the 
order $\lambda^3$ and beyond does not seem to make much sense. 

To conclude this section let us remark that the weak coupling
expansions (both the low-energy and semiclassical) of the meson masses 
are only asymptotic ones. For the
expansions of the BS eigenvalues ${\widetilde M}_n$ this fact can be deduced
from their analytic structure as the functions of $\lambda$: there are
infinitely many complex singularities accumulating at the point
$\lambda=0$, as we explain in Sect.9. It is possible to argue that the 
point $\lambda=0$ is an essential singularity of the actual masses
$M_n$ as well, but this discussion goes beyond the scope of the
present paper.

\section{Multi-quark corrections}

As was already mentioned in the Section 4, the Bethe-Salpeter equation
ignores substantial part of the multi-quark contributions to the meson
masses. In general, explicit computation of such contributions is 
a rather difficult task which we do not attempt to perform
here. Instead, we briefly discuss the role of the multi-quark
corrections, in particular the way they modify the weak-coupling
expansions described in the previous Section.

The multiquark corrections are represented by diagrams
with more then two quarks in the intermediate states, like the diagram in
Fig.3b. Formally, all such terms can be considered as the higher-order
corrections to the kernel in the right-hand side of the
equation \eqref{inteq}, 
\bea\label{modker}
{G}_P (p|q) \ \to\ {\widetilde G}_P (p|q) = \ {
G}_{P}^{(0)}(p|q) + \lambda\,{G}_{P}^{(2)} (p|q) + 
\lambda^2\,{G}_{P}^{(3)}(p|q) +
\cdots\ ,
\eea 
where $G_{P}^{(0)}(p|q) \equiv G_{P}(p|q)$ is the original kernel
\eqref{gpgg}. All terms here are expected to be singular at $p=\pm
\,q$, the singularities being of two types: the delta-function
singularities coming from the disconnected diagrams (like the one
shown in Fig.3c), and the second-order poles at $p\pm q =0$ reflecting
the long-range character of the interaction in \eqref{hift}. Let us
discuss first the role of these singular terms.

The delta-function terms were previously discussed in the Section
4. Their role is to modify the quark self-energy $\omega(p) \to 
\varepsilon(p)$, and to renormalize the quark mass. The 
first part is already taken into account in the left-hand side of 
the equation \eqref{inteqo}\,, and since the associated one-particle 
reducible diagrams disappear in the limit $P=\infty$, this part plays
no role in the final form of the Bethe-Salpeter equation \eqref{bs}.
The second part - the quark-mass corrections due to the one-particle 
irreducible diagrams - can be taken into account if one replaces the 
parameter $m$ in the Eq.\eqref{bs} by the full ``dressed'' quark mass
$m_q$. The latter is defined perturbatively, as the sum of all
one-particle irreducible self-energy diagrams. By this definition the 
quark mass is given as the power series
\bea\label{mqseries}
m_{q}^2 = m^2\,\big(1 + a_2\,\lambda^2 + a_3\,\lambda^3 + \cdots\big)\ .
\eea
In principle, the coefficients $a_k$ can be determined through well
defined calculations within the perturbation theory. However, the
problem is rather involved, and at the moment only the first
correction term in \eqref{mqseries} is known exactly \cite{fz2},
\bea\label{mqtwo}
a_2 = 0.071010809...\ .
\eea
This and the higher order corrections in \eqref{mqseries} obviously
modify the weak coupling expansions of the meson masses $M_{n}$. Thus, 
in the low-energy expansions of the actual meson masses
\bea\label{lowactual}
{{M^2}\over{4\,m^2}} = 1 + \sum_{k=1}^{\infty}\,\mu_k \,t^k
\eea
the coefficients $\mu_k$ coincide with the corresponding coefficients 
$\epsilon_k$ in \eqref{lowexp} at $k <6$, but starting from $\mu_6$ they
receive additional contributions due to the quark mass corrections. 
In particular, at the order $t^6$ we have
\bea\label{musix}
\mu_6 = \epsilon_6 + a_2\,, 
\eea
while the next term $\sim t^7$ receives no correction, so that we
still have $\mu_7 =0$. 
Yet higher orders are affected by these as well as by other types 
of the multi-quark corrections, as we explain below. Our present
knowledge about the ``dressed'' quark mass $m_q$ as the function of 
$\lambda$ is very limited. No higher order coefficients $a_3, a_4, \ldots$ in
\eqref{mqseries} are known even approximately. Moreover, there are no
reason to expect the perturbative expansion \eqref{mqseries} to
converge. On a more fundamental level, no clear definition of the
dressed quark mass beyond the perturbation theory is
available. Therefore, apart from the weak coupling expansions, the
quark mass $m_q$ can be treated presently as a phenomenological
parameter at best. Taking this point of view we will observe in
the Section 8 that the overall radiative correction 
$\delta m \equiv m_q - m$ is probably very small
(no more then $\sim 2-3\%$ of the lightest meson mass $M_1$) at all
positive $\lambda$.

Next, the even order terms in \eqref{modker} are expected to have
the second order poles at $p\pm q =0$. The pole terms are
interpreted as the corrections to the string tension: the
``bare'' string tension $f_0 = 2{\bar\sigma}\,h$ in the r.h.s. of the
BS  equation \eqref{bs} is effectively replaced by the ``dressed'' one
$f=f(h)$. To make it precise, we write
\bea\label{f0f}
f_{0}\,{\widetilde G}_{P}(p|q) = f\,{\mathbb G}_P (p|q)
\eea
where by definition the kernel ${\mathbb G}_P$ is normalized in such a
way that its second order poles have the residues $1$, as in 
\eqref{doublepoles}; we can write
\bea\label{newker}
{\mathbb G}_P (p|q) = G_P (p|q) + 
\Delta {\mathbb G}_{P}^{\rm (reg)}(p|q)
\eea
where $G_P (p|q)$ is the original kernel \eqref{gpgg} which has the
standard pole terms \eqref{doublepoles}, and the last term is regular 
at $p=\pm\,q$ \footnote{Although regular
at $p\pm q =0$, the term $\Delta {\mathbb G}_{P}^{\rm (reg)}(p|q)$ certainly
has other singularities; even when $p$ and $q$ are real one expects to have
branch cuts associated with the multi-particle thresholds.}. In this way
the coefficient $f$ appears as a power series in $\lambda^2$,
\bea\label{fullf}
f = f_0\,\big(1 + c_2\,\lambda^2 + \cdots\big).
\eea
It is possible to argue that the coefficients $c_{2k}$ here are
related in a simple way to the coefficients ${\tilde g}_{2k+1}$ of the 
weak field expansion
\bea\label{fser}
{{{\cal F}_{\rm vac}}\over {m^2}} - {{1}\over{8\pi}}\,\log m^2  = 
-\,{1\over 2}\,\lambda + {\tilde g}_2\,\,\lambda^2 + {\tilde
  g}_3\,\,\lambda^3 + {\tilde g}_4\,\,\lambda^4 + \ldots\ ,
\eea
of the IFT vacuum energy density ${\cal F}_{\rm vac}\,$ 
\footnote{The vacuum energy density ${\cal F}_{\rm vac}\,\,$ is of
  course the same as the (singular part of) specific free energy of 
the near-critical Ising model.}. Let us write ${\cal F}(h)$ for the 
analytic function of $h$ which coincides with
${\cal F}_{\rm vac}$ at {\it positive} real $h$. Then the analytic 
continuation of ${\cal F}(h)$ to the negative real values of $h$
describes the energy density ${\cal F}_{\rm meta}$ of the ``false'' 
vacuum -- the Lorentz invariant metastable state having the
magnetization opposite to the external field $h$ (e.g. $\langle
\,\sigma(x)\,\rangle_{\rm meta} < 0$ at $h > 0$). It is natural to 
interpret the difference 
\bea\label{deltaf}
{\cal F}_{\rm meta} - {\cal F}_{\rm vac}
\eea
as the effective string tension. Indeed, when the 
two quarks constituting a meson are widely separated (the
configurations responsible for the singular terms in \eqref{newker}), 
the spatial domain between the quarks is effectively filled with the 
``false'' vacuum. It follows that the coefficients in \eqref{fullf}
are expressed through ${\tilde g}_n$ as
\bea\label{ctog}
c_{2k} = -2\,{\tilde g}_{2k+1}
\eea
This relation provides several terms in \eqref{fullf}, since many
coefficients in \eqref{fser} are known \footnote{The related
coefficients ${\tilde G}_n = (2{\bar s})^n\,\,{\tilde g}_n$ are
collected in Table 2 of the Ref.\cite{fz1} . }. In particular
\bea\label{ctwo}
c_2 = -0.003889...
\eea

The correction terms
in \eqref{fullf} also make contributions to the low energy expansion
\eqref{lowactual}. The first term affected is $\mu_8\,t^8\,$,
\bea\label{mueight}
\mu_8 = \epsilon_8 + {{2\,c_2 + a_2}\over 3}\,\epsilon_2\,.
\eea
Corrections to the terms $\sim t^9$ and higher involve additional 
contributions from the regular term in \eqref{newker}, see below.

It is tempting to take the difference \eqref{deltaf} as the
nonperturbative definition of the effective string tension $f$. 
It cannot be valid literally though. The function ${\cal F}(h)$ is
known to have a branch cut along the negative part of the real axis
\cite{nucleation}, hence the analytic continuation defining 
${\cal F}_{\rm meta}$ in fact returns complex values. The imaginary 
part is interpreted in terms of the tunneling decay probability
\cite{volosha} and, being exponentially small in $h$, it is invisible in the 
perturbative expansion \eqref{fullf}. In fact, both ${\cal F}_{\rm
  vac}$ and ${\cal F}_{\rm meta}$ are known numerically with high
precision, see Ref.\cite{fz1}. 
\begin{figure}[ht]
\centering
\includegraphics[width=14.0cm]{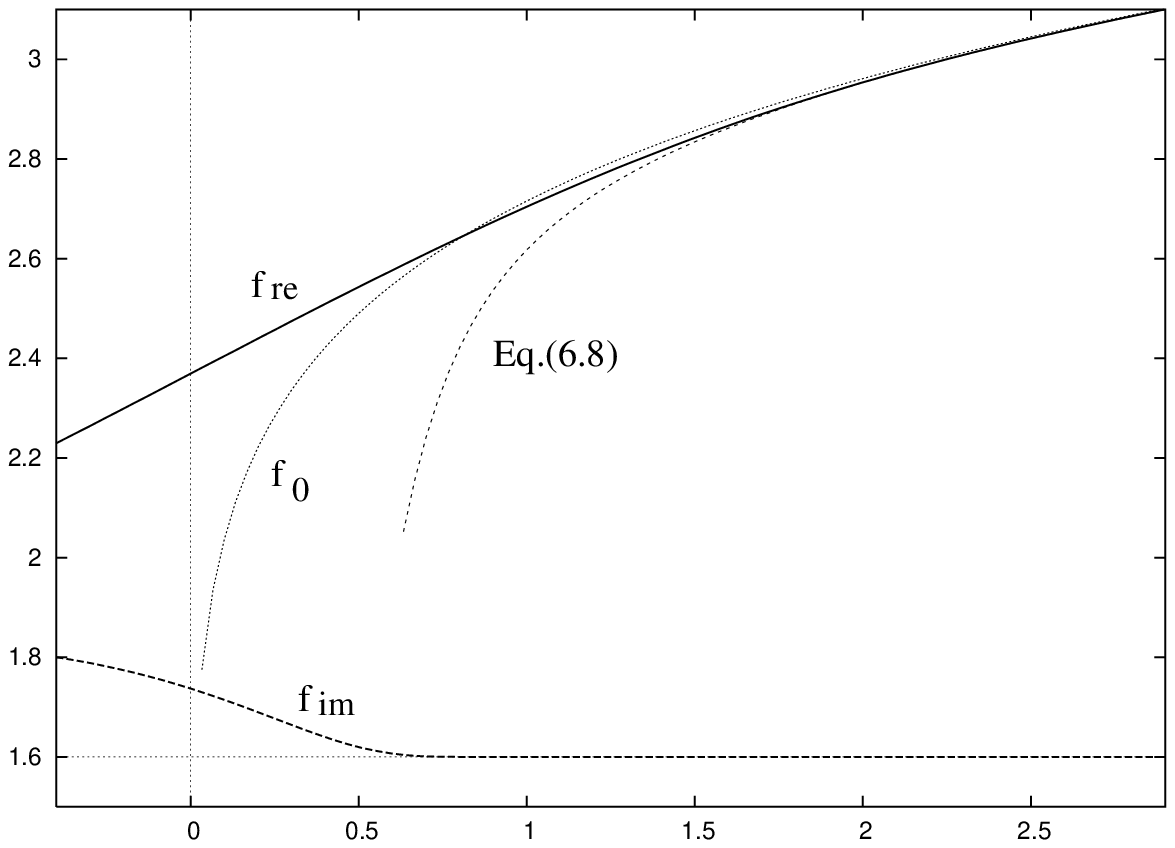}
\caption{\small Plot of the difference \eqref{deltaf} measured in the units of
$|h|^{16\over 15}$, vs the scaling parameter $\eta$. The solid line
and the
``fat''\,dashed line represent the real and the imaginary parts 
$f_{\rm re} = \Re e\,[{\cal F}_{\rm meta} - {\cal F}_{\rm vac}]$ and 
$f_{\rm im} = \Im m\,({\cal F}_{\rm meta} - {\cal F}_{\rm vac})$, 
respectively. The imaginary part is shifted upward by $1.6$ to make
the Figure more compact. The dotted line represents the bare string tension 
$f_0 = (2{\bar s}\,\eta^{1\over 8})\,|h|^{16\over 15}$, and the fine dashed
line shows the effect of the first correction to $f_0$ explicitly
written in Eq.\eqref{fullf}.}
\label{fig-deltaf}
\end{figure}
In Fig.6 we show a plot of the difference
\eqref{deltaf} (measured in the units of $|h|^{16\over 15}$) as the
function of the scaling parameter \eqref{etadef}. One can see that 
the imaginary part of \eqref{deltaf} is much smaller then the real
part at all positive $\eta$ (moreover, it essentially vanishes at
$\eta \gtrsim 0.8$), therefore taking the real part seems to provide a
reasonable definition for the effective string tension $f$.
Note that the real part differs very little from the bare string
tension $f_0$ at all but very small (positive) $\eta$. The difference
is significant at $\eta \lesssim 0.8$, and becomes drastic at $\eta \to
0$. Indeed, at small $\eta$ the bare tension tends to zero in a
singular way, $f_0 \sim \eta^{1\over 8}$, while the difference
\eqref{deltaf} has finite limit at $\eta=0$, and is analytic in some
domain around this point. In fact, a number of terms of its Taylor
expansion around $\eta=0$ can be extracted from the results of
\cite{fz1}; for instance, for the real part $f_{\rm re}$ of
\eqref{deltaf} we have
\bea\label{deltafser}
f_{\rm re} = |h|^{16\over 15}\,\big(\rho_0 + \rho_1\,\eta +
\rho_2\, \eta^2 + \cdots\big)\ ,
\eea
where the coefficients are expressed through $\Phi_k$ defined in 
Ref.\cite{fz1} (see Table 3 there) as 
$\rho_k = \big(\cos{{8\pi\,(k-2)}\over 15}-1\big)\,\Phi_k\,$, for
instance
\bea\label{rhoo}
\rho_0 = 2.3692934...\,, \qquad \rho_1 = 0.3521342...\,, \qquad \rho_2
=0\,, \qquad {\rm etc}
\eea

According to the above analysis, the singular parts of the multi-quark
corrections in \eqref{modker} modify the weak coupling expansions 
in a relatively trivial way, through the renormalizations of the 
parameters $m$ and $f_0$ entering the BS equation \eqref{bs}, which
are to be replaced by the ``dressed'' parameters $m_q$ and $f$. Thus,
the singular terms are taken into account by substituting
\bea\label{subsmf}
m \rightarrow m_q\ , \qquad \lambda \rightarrow {f/{m_{q}^2}}
\eea
in the low-energy expansion \eqref{lowseries}, and the same
applies to the semiclassical expansion \eqref{bsccc} as
well. Remaining multi-quark corrections are associated with the
regular term in \eqref{newker}. It is not difficult to show that 
this kind of corrections affects the low-energy expansion
\eqref{lowactual} only at the order $t^9$ and higher. Indeed, 
the term $\Delta {\mathbb G}_{P}^{(reg)}(p|q)$ is regular at
sufficiently small $p$ and $q$, and it is an odd function of each of
this variables. Therefore its leading low-energy behavior is
\bea\label{regkerlow}
\Delta{\mathbb G}_{P}^{({\rm reg})}(p|q) \to A(P)\,\,p\,q + \cdots\ .
\eea
Since this term is there entirely due to the higher-order corrections,
the coefficient $A(P)$ here is $\sim \lambda$. Recall also that the 
low-energy domain corresponds to $p,q \sim t$. It follows that the
correction to the BS equation due to this regular term is $\sim
t^9 = \lambda^3$. Similarly, one can check that the regular term leads to
the corrections to the semiclassical quantization condition
\eqref{bsccc} at the order $\sim \lambda^3$ and higher. Therefore
developing the weak coupling expansions (both the low-energy and the 
semiclassical ones) of the actual meson masses at the order
$\lambda^3$ and beyond requires explicit evaluation of the contributions
from the ``regular'' term $\Delta {\mathbb G}_{P}^{\rm (reg)}$ in
\eqref{newker}. The latter constitutes a separate and rather involved
problem which we do not attempt to handle here. Note that the
correction term $\sim \lambda^3$ to the quark mass in the
Eq.\eqref{mqseries}, whose coefficient $a_3$ is also presently
unknown, contributes at the same order. To summarize, existing data are
sufficient to determine the coefficients of the weak coupling
expansions of $M_n$ all the way up to, but not including, the order
$\lambda^3$. Evaluating further terms requires additional ingredients,
most importantly developing a perturbative expansion of the term $\Delta
{\mathbb G}_{P}^{\rm (reg)}$ and the higher-order corrections to the
quark mass. 

The above analysis, in particular the low-energy expansion, apply to
stable mesons, with the masses $M_n$ well below the stability
threshold $2 M_1$, twice the mass $M_1$ of the lightest meson. As 
given meson mass approaches to the threshold, one expects the
threshold singularities of the last term in \eqref{newker} to play
more and more imoportant role. As the result, when close to the
threshold the actual mass $M_n$ is expected to deviate significantly 
from its BS approximation ${\widetilde M}_n$. Furthermore, when the 
meson mass exceeds the threshold the actual mesons turn into resonances
(see footnote$^6$ on page 4), and their masses $M_n$ develop
imaginary parts representing the decay probabilities. This
effect is certainly outside the scope of the two-quark approximation:
the BS equation \eqref{bs} describes an infinite tower of stable
mesons. And even including the multi-quark states perturbatively, as
was discussed above, does not seem to provide a consistent way of
treating the mesons above the stability threshold. Let us mention 
interesting attempt to treat the issue of the unstable mesons on the
perturbative level in Ref.\cite{rut}. However, we believe that
systematic treatment of the decay problem must involve at least some
understanding of the meson scattering states, the problem which we
hope to address in the future.

\section{Numerical solution} 

As we will argue in the next Section, the usefulness of the
Bethe-Salpeter equation goes well beyond the derivation of the
weak-coupling expansions \eqref{lowseries}, \eqref{bsccc}\,. Therefore  
it is of interest to study the eigenvalues ${\widetilde M}_{n}^2$ of the BS
equations \eqref{bsh} at all values of $\lambda$, real and complex 
\footnote{There are good reasons to expect that the BS masses
${\widetilde M}_n$, taken as the functions of complex $\lambda$
(or rather of the complex variable $\bar\eta = 1/\sqrt{\lambda}$),
imitate much of the analytic properties of the actual masses $M_n$ in
the complex $\eta$\,-\,plane. We will say more on this subject at the
end of this Section, and again in Sect.9. In our view, gaining any
insight into the analytic properties of $M_n(\eta)$ (structure of its
Riemann surface, singularities, etc) is of central importance for
overall understanding of IFT. This problem was one of primary
motivations of our work.}. The equation \eqref{bs} is unlikely to
admit an analytic 
solution, but its numerical solution is not difficult to develop. The 
only feature which makes the numerical approach not exactly
straightforward is the presence of the singularities of the
kernel \eqref{kernelt} at $\theta' = \pm\,\theta$. This is one of 
many features that the equation \eqref{bs} shares with the mass
spectrum equation in the 't Hooft's model \cite{thooft}. Several 
methods have been developed for numerical  solution of the latter 
(see \cite{thooft,matthias}). We found it convenient to
use yet another approach which is based on the Fourier-transformed
version of the equation \eqref{bs}. We introduce the rapidity
Fourier-transform 
\bea\label{psitilde}
\Psi(\theta) =
\int_{-\infty}^{\infty}\,{\psi}(\nu)\,e^{-i\nu\theta}\,d\nu\ , \qquad
{\psi}(\nu) =
\int_{-\infty}^{\infty}\,{\Psi}(\theta)\,e^{i\nu\theta}\,
{{d\nu}\over{2\pi}}\ .
\eea
In terms of ${\psi}(\nu)$ the norm \eqref{metric} becomes
\bea\label{nunorm}
||\Psi||^2 = {1\over
  4}\,\int_{-\infty}^{\infty}\,{\psi}^{*}(\nu)
  K(\nu-\nu'){\psi}(\nu') \,d\nu\,d\nu'\ ,
\eea
where
\bea\label{kdef}
K(\nu)\ =\
\int_{-\infty}^{\infty}\,{{e^{i\nu\theta}}\over{\cosh^2\theta}}\,
{{d\theta}\over{2\pi}}\ = \ {\nu\over{2\,\sinh{{\pi\nu}\over
      2}}}\ ,
\eea
and then the equation \eqref{bs} transforms to
\bea\label{bsnu}
8\,\bigg(m^2 + f_0\,\,\nu \tanh{{\pi\nu}\over
  2}\bigg){\psi}(\nu) - {{f_0}\over
  2}\,{\nu\over{\cosh{{\pi\nu}\over
      2}}}\,\int_{-\infty}^{\infty}\,{{\nu'}\over{\cosh{{\pi\nu'}\over
      2}}}\,{\psi}(\nu')\,d\nu' = \nonumber\\
{{M^2}}\,\int_{-\infty}^{\infty}\,[K(\nu-\nu')-K(\nu+\nu')]\,
{\psi}(\nu')\,d\nu'
\eea
Now the kernel is regular at any real arguments, and the equation
\eqref{bsnu} admits numerical solution through straightforward
discretization of the variable $\nu$. Before turning to the numerics,
let us say few words about the general properties of the BS equation
written in this form.

It is not difficult to show that generic solution ${\psi}(\nu)$ has
poles at the values of $\nu$ which solve the equation
\bea\label{nueq}
1 + \lambda\,\nu\,\tanh{{\pi\nu}\over 2} =0\ .
\eea
where again $\lambda = f_0/m^2$. At real positive $\lambda$ all
solutions are purely imaginary, of the form $\pm\,i\,\kappa$, $\kappa$
being a positive root of the equation
\bea\label{kappaeq}
\lambda\,\kappa\,\tan{{\pi\kappa}\over 2} = 1\ .
\eea
Let $\kappa_0$ be the lowest of such roots; at real positive
$\lambda$ it lays in the interval $[0,1]$. The associated pole of
$\psi(\nu)$ controls the large\,-\,$\theta$ behavior of the
rapidity-space wave function,
\bea\label{psilarge}
\Psi(\theta) \to r\,\,{\rm sign}(\theta)
\,\,e^{-\kappa_0\,|\theta|} \qquad {\rm as} \quad |\,\theta\,|
\to \infty\,.
\eea
where $r$ is a (normalization-dependent) constant. Note that at large
$\lambda$ this pole approaches the real axis, since
\bea\label{kappal}
\kappa_0 \to \sqrt{2\over{\pi\lambda}}\qquad {\rm as}\quad \lambda\to
\infty\ .
\eea
At $\lambda=\infty$ every eigenfunction $\psi_n (\nu)$ has a
simple pole at $\nu=0$; correspondingly, the associated rapidity wave
function $\Psi_n(\theta)$ tends to a constant $r_n$ at large $|\theta|$. 
One consequence of this phenomenon is the nature of the expansion
around the point $\lambda = \infty$. This point appears to be a
square-root branching point of the eigenvalues ${\widetilde M}_{n}^2$ taken 
as the functions of $\lambda$ (see remark at the end of this
Section). Therefore the large\,-\,$\lambda$ expansions of ${\widetilde
  M}_{n}^2$ are of the form  
\bea\label{yseries}
{{{\widetilde M}_{n}^2}\over {4\,m^2}} = Y_{n}^{(0)}\,\lambda +
Y_{n}^{(1)}\,\sqrt{\lambda} + Y_{n}^{(2)} + ... +
Y_{n}^{(k)}\,\lambda^{1-{k\over 2}} + ...\ ,
\eea
where $Y_{n}^{(k)}$ are constants, and the series converge in finite
domains. It is straightforward to derive the following expression for
the coefficients at the terms $\sim \sqrt{\lambda}$\,,  
\bea\label{yone}
Y_{n}^{(1)} =
{1\over{2\,\sqrt{2\pi}}}\,\bigg[{{|r_{n}|^2}\over{||\Psi_{n}||^2}}
\bigg]_{\lambda=\infty}\ ,
\eea
where $r_n$ are the constants in the asymptotics \eqref{psilarge} of
the wave functions $\Psi_n$.

\begin{table}[ht]
\begin{center}
\begin{tabular}{| r | c | l | l | l | l | l |}
\hline
\rule{0mm}{4mm}
$n$\hspace{1mm}
& $\lambda = 0.01$ & $\ \ \lambda = 0.10$ & $\ \ \lambda =
1.00$ & $\ \ \lambda = 10.0$ & $\ \ \lambda = 100$ \\
\hline
\rule{0mm}{3.6mm}
1  & 1.110840242 & 1.550883388 & 4.185654285 & 24.29578962 & 205.1229687\\
2  & 1.196814699 & 2.024246733 & 7.637285443 & 56.34010255 & 523.5675928\\
3  & 1.269079423 & 2.446855032 & 10.96393850 & 88.01258645 & 838.4720639\\
4  & 1.334331298 & 2.845183371 & 14.23862629 & 119.5937889 & 1152.960249\\
5  & 1.395100728 & 3.228787972 & 17.48445168 & 151.1324905 & 1467.318135\\
6  & 1.452678928 & 3.602360894 & 20.71182429 & 182.6463568 & 1781.617148\\
7  & 1.507833369 & 3.968598364 & 23.92635268 & 214.1438592 & 2095.883687\\
8  & 1.561065333 & 4.329208970 & 27.13141319 & 245.6299865 & 2410.129928\\
9  & 1.612724086 & 4.685350869 & 30.32919948 & 277.1195629 & 2724.362364\\
10 & 1.663064726 & 5.037848886 & 33.52121875 & 308.8383150 & 3038.584826\\
\hline
\end{tabular}
\end{center}
\caption{\small Ten lowest eigenvalues of the BS equation
      \eqref{bsh} at different values of $\lambda$ obtained by
      numerical solution of the Eq.\eqref{bsnu}. The ratios
      ${\widetilde M}_{n}^2/4\,m^2$ are presented.}
\label{table-ampl}
\end{table}

As was already said, numerical solution of \eqref{bsnu} is obtained
by discretization of the variable $\nu$. In Table 1 we 
present few lowest eigenvalues ${\widetilde M}_{n}^2$ for some values of
$\lambda$. At substantially smaller $\lambda$ ($\lambda \leq 0.001$)
the numerical results are indistinguishable (within the given accuracy)
from the first eight terms of the low-energy expansion
\eqref{lowseries}. On the other side, at large $\lambda$, the
expansions \eqref{yseries} apply. Numerical estimates of the three leading 
coefficients $Y$ for few lowest mesons are given in the Table 2.
It is interesting to observe that as $n$ grows the leading coefficients
$Y_{n}^{(0)}$  quickly approach the following simple asymptotic form
\bea\label{yass}
Y_{n}^{(0)} \ \to \ \pi\,(n-3/8)\qquad {\rm as}\quad n \to \infty \ .
\eea
In the same limit $Y_{n}^{(1)}$ tend to a constant value $1.2533...$,
while $Y_{n}^{(2)}$ increases logarithmically as ${1\over 2}\log (n-3/8) + 
{\rm Const}$, with Const $\approx 1.209$. 

 \begin{table}[ht]
\begin{center}
\begin{tabular}{| r | l | l | l | r | l | l | l |}
\hline
\rule{0mm}{4mm}
$n$ & $\ \ \ \ \ Y_{n}^{(0)}$ & $\ \ \ \ \ Y_{n}^{(1)}$ & $
\ Y_{n}^{(2)}$ & $n$ & $\ \ \ \ \ Y_{n}^{(0)}$ & $\ \ \ \ \ Y_{n}^{(1)}$ 
& \ $Y_{n}^{(2)}$ \\ 
\hline
\rule{0mm}{3.6mm}
1  & 1.904347602  & 1.3810744  & 0.873 & 6  & 17.67032120  &
1.25432903  & 2.072 \\
2  & 5.094667187  & 1.2671521  & 1.450 & 7  & 20.81220388  &
1.25403891  & 2.154 \\ 
3  & 8.242239347  & 1.2582803  & 1.691 & 8  & 23.95398627  &
1.25385735  & 2.224 \\ 
4  & 11.38577501  & 1.2558373  & 1.853 & 9  & 27.09571021  &
1.25373628  & 2.286 \\  
5  & 14.52825003  & 1.2548349  & 1.974 & 10 & 30.23739767  &
1.25365155  & 2.341 \\  
\hline
\end{tabular}
\end{center}
\caption{\small Three leading coefficients of the expansion \eqref{yseries} for
the first ten eigenvalues ${\widetilde M}_{n}^2$. The Eq.\eqref{yone} was
used in computing $Y_{n}^{(1)}$.}

\label{yns}
\end{table}

Numerical diagonalization can be applied also at complex
$\lambda$. The procedure is straightforward as long as $|{\rm
  arg}\,\lambda\,| < \pi$. But at ${\rm arg}\,\lambda = \pm\,\pi$ the
poles $\nu=\pm\,i\kappa_0$ surface at the real $\nu$\,-\,axis, and if one
continues to the next sheet $|{\rm arg}\,\lambda\,| >\pi$ these poles break
through the real axis, leaving behind the residue terms. Thus, at the
second sheet one has to consider the equation which differs from the 
Eq.\eqref{bsnu} by this residue term. Continuing around the point
$\lambda=\infty$ second time generates another residue term which
cancels exactly the first one. It follows that $\lambda=\infty$ is
the square-root branching point of the solution, as was stated in the
Eq.\eqref{yseries} above. Furthermore, one can argue that the
BS masses ${\widetilde M}_n$, taken as the functions of complex
$\lambda$ exhibit infinitely many singularities on
the second sheet of the $\lambda$\,-\,plane. Indeed, as was explained
above, when continuing to the second sheet ${\rm arg}\,\lambda > \pi$,
the pole at $\nu = i\kappa_0$ sinks into the lower half-plane, where
at certain values of $\lambda$ it collides with other poles of
$\psi(\nu)$, associated with the higher roots of the equation
\eqref{kappaeq}. It is possible to show that these
singularities accumulate towards the point $\lambda=0$, making this
point an essential singularity of all BS masses. This picture was
behind our statement in Section 5 about the asymptotic nature of
the weak coupling expansions. But detailed understanding of the
analytic properties of the BS masses ${\widetilde M}_n$ as the
functions of complex $\lambda$ remains to be achieved, and we intend
to come back to this problem elsewhere.

\section{Comparison to the TFFSA data}

By construction, the BS equation was designed to describe stable
mesons at sufficiently small $\lambda$, where the idea of the
mesons as predominantly two-quark constructs is well
justified. However, it turns out that this approximation describes 
the mass spectrum of stable mesons remarkably well for all but very
large values of $\lambda$. Moreover, when some multi-quark effects
(notably the renormalizations of the string tension) are taken into
account, the approximation works very well at all real positive
$\lambda$. In this Section we compare the results derived from the BS 
equation with the data on the mass spectrum obtained directly from the 
IFT using the TFFSA\cite{fz1}. It is convenient in this context to
discuss in terms of the scaling parameter \eqref{etadef} related to 
$\lambda$ as
\bea\label{lambdaeta}
\lambda = {{2{\bar\sigma}h}\over m^2} = {{2\,{\bar
      s}}\over{\eta^{15\over 8}}}\ ,
\eea
where ${\bar s}$ is the constant defined in \eqref{sexact}\,. In 
the following discussion we still reserve the notation $M_n$ (or $M_n
(\eta)$ when the dependence on $\eta$ is to be emphasized) for the
actual masses of the particles in the IFT, and use the notation
${\widetilde  M}_n$ for the BS masses, obtained from the equation
\eqref{bs}. In fact, we will often write ${\widetilde M}_n (m, f_0)$
to signify their dependence on the parameters in the Eq.\eqref{bs}.

We do not describe here the TFFSA in any details, which are presented 
in Ref.\cite{fz1}\,. The strategy for extraction the particle
masses is not much different from that commonly used in the TCSA
\cite{tcsa}\,. We have employed additional improvements (mostly
in the way the finite-size effects are taken into account) which
will be described in due details in a separate paper. We have used this
technique to determine the actual masses $M_n$ of stable particles, 
at positive $\eta$ ranging from $0$ to $24$. With the truncated space
involving approximately 1800 lowest levels we were able to achieve
typical accuracy of five significant digits for all those values of
$\eta$, except for when the mass $M_n$ is very close to the stability
threshold $2\,M_1$, where the accuracy may be somewhat lower. The results
were shown in Fig.1 for $\eta\in[-1:4]$ (at greater values of $\eta$
deviations from the weak coupling approximations \eqref{lowseries} or
\eqref{bsccc} would not be visible). Beyond the threshold $2 M_1$ the
particles become resonance states, their masses $M_n$ acquiring
imaginary parts\,\footnote{It is possible
to extract the masses and the widths of the resonances from the TFFSA
data, and we have preliminary results for few lowest resonances.
However, so far precision of our data is far from being satisfactory,
especially when the resonances are wide, and we refrain from reporting
this data here.}. At $\eta=0$ most of the resonances disappear,
while some become stable particles again, in accord with the spectrum of the
integrable IFT at $\eta=0$. The masses at $\eta=0$ are known exactly 
\cite{me1,fatey}, and TFFSA reproduces them within the stated accuracy. For
all we know, the integrable point $\eta=0$ is analytic, in particular
the masses $M_n = M_{n}(\eta)$ admit power series expansions 
\footnote{Shifting away from the integrable point renders unstable all
but the three lowest particles; correspondingly, the coefficients
$M_{n}^{(2)}$ (and higher) with $n>3$ generally have nonzero imaginary parts 
\cite{delfino2,res1}.} 
\bea\label{mnseries}
M_{n}(\eta) = M_{n}^{(0)} +
M_{n}^{(1)}\,\eta + M_{n}^{(2)} \,\eta^2 + ...\ .
\eea
The slopes $M_{n}^{(1)}$ for few lowest $n$ are also known exactly,
through the form-factor perturbation theory \cite{delfino2}, and our
TFFSA data is in full accord with the exact results (for previous
numerical verification of the exact slopes using the lattice model see
Ref.\cite{slopes}). These numbers, as well as numerical estimates of the 
next term coefficients for the lowest mesons are collected in the
Table 3.
\begin{table}[ht]
\begin{center}
\begin{tabular}{| r | r | r | r | r | r | r |}
\hline
\rule{0mm}{4mm}
$n$ & $M_{n}^{(0)}\ \ \ \ \ \ \ $ & $M_{n}^{(1)}\ \ $ & \
$M_{n}^{(2)}\ \ $ & ${\widetilde M}_{n}^{(0)}\ \ \ \ \ $ & ${\widetilde
  M}_{n}^{(1)}\ $ & \ ${\widetilde M}_{n}^{(2)}$ \\ 
\hline
\rule{0mm}{3.6mm}
1  &\ 4.404908579981 & 1.295045 &  0.2003 &  4.248274145 & 1.316490 & 0.2815 \\
2  &\ 7.127291799746 & 1.115886 &  0.2072 &  6.948600321 & 1.077762 & 0.3726 \\ 
3  &\ 8.761556059774 & 1.953268 & -1.5110 &  8.838163544 & 1.095065 & 0.3471 \\ 
4  &\ 10.59322004129 & 1.484334 & ****** & 10.387731645 & 1.144113 & 0.3212 \\  
5  &\ 13.02221009790 & ******* & ****** & 11.733999773 & 1.253651 & *****\\  
\hline
\end{tabular}
\end{center}
\caption{\small The coefficients of the Taylor expansions
\eqref{mnseries} for four lowest mesons. The leading coefficients
$M_{n}^{(0)}$ are exact 
masses of the integrable theory at $\eta=0$ \cite{me1,fatey}. The
exact slopes $M_{n}^{(1)}$ are taken from Ref.\cite{delfino2}, while
the coefficients
$M_{n}^{(2)}$ are estimated from the TFFSA data. The BS coefficients
${\widetilde M}_{n}^{(k)}$ are computed using \eqref{yseries} with
$\lambda=f/m^2$, and the expansion \eqref{deltafser} for $f$.}
\label{mns}
\end{table}

As expected, at large $\eta$ the masses from TFFSA agree very well
with the low-energy expansions \eqref{lowactual}. This is demonstrated
in Fig.7,  
\begin{figure}[ht]
\centering
\includegraphics[width=14cm]{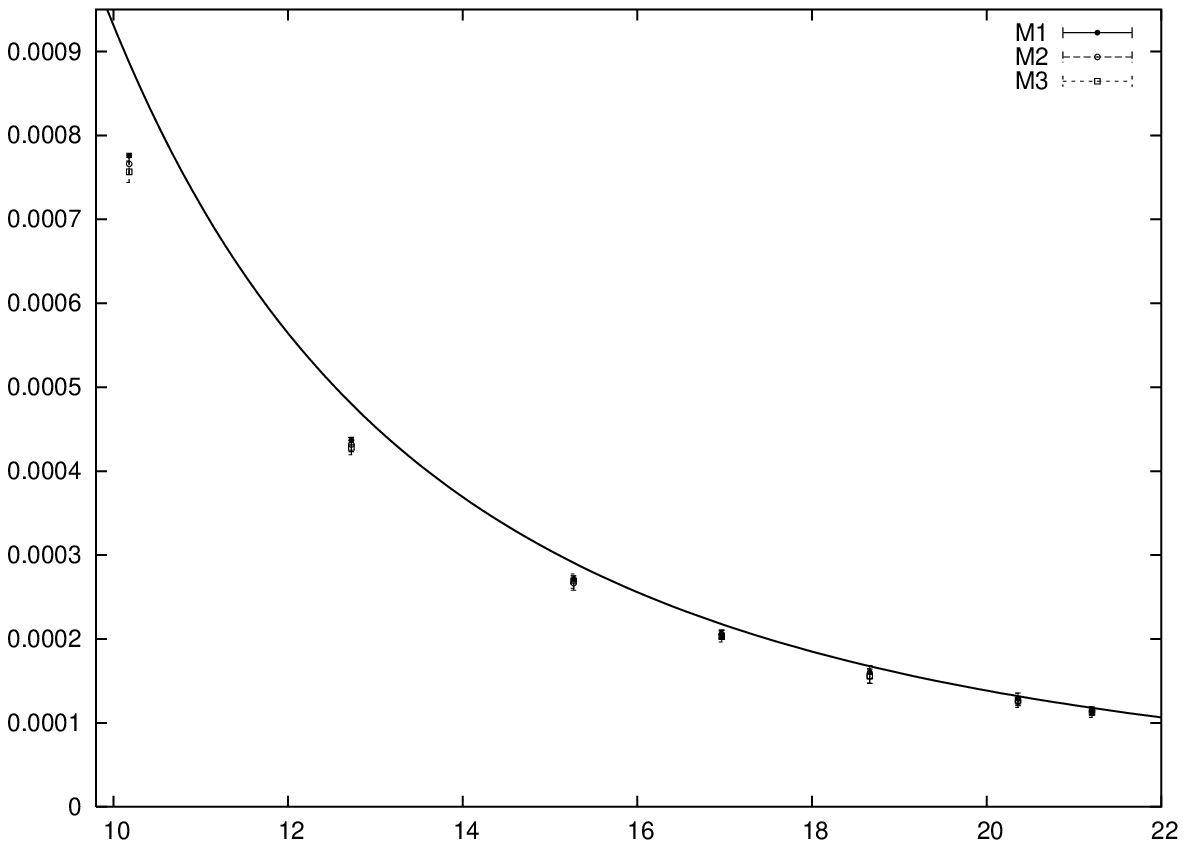}
\caption{\small Plots of the differences $M_n - 
{\widetilde M}_{n}(m,f_0)$ (in the
  units of $|h|^{8\over 15}$), vs the scaling parameter $\eta$. The
  solid line is the curve $(4\,a_2\,{\bar
    s}^2)\,\eta^{-11/4}$ representing the correction $m\,a_2\,t^6$
  from the term $a_2$ in Eq.\eqref{musix}.}
\label{fig-deltam1}
\end{figure}
where the differences $M_n (\eta) - {\widetilde M}_n
(m, f_0)$ (with $\eta = (2{\bar s} m^2/f_0)^{8/15}$) for the three
lowest mesons are shown at large $\eta > 10$. In this domain
the BS masses ${\widetilde M}_n (m,f_0)$ are accurately described by the
low-energy expansions \eqref{lowseries}. According to the arguments in
Sect.6, the most part of the differences must be attributable to the
quark mass renormalization \eqref{mqseries}, since its effect on the
meson masses is $\sim\lambda^2$, while other multi-quark
corrections are of higher orders. At large $\eta$ the deviations
look consistent with the asymptotic form $(a_2\,\lambda^3)\,m = 
(4\,a_2\,{\bar s}^2\,|h|^{8\over 15})\,\eta^{-11/4}$ which corresponds
to the leading quark mass correction (see Eq.\eqref{musix}). The
remaining mismatch visible in Fig.7 is likely to come from the
unaccounted (and presently unknown) higher-order corrections to the
quark mass, and perhaps from the multi-quark corrections to the
regular part of the effective kernel discussed at the end of the
Sect.6.  

It turns out that even if all renormalization effects are disregarded,
the BS masses ${\widetilde M}_{n}(m,f_0)$ approximate the actual
masses $M_n$ very well at all but very small $\eta$. As is seen in the 
plot in Fig.8, the BS masses ${\widetilde M}_{n}(m,f_0)$ significantly 
deviate from $M_n$ only at $\eta\lesssim 0.5$. Note that this is
\begin{figure}[t]
\centering
\includegraphics[width=14.0cm]{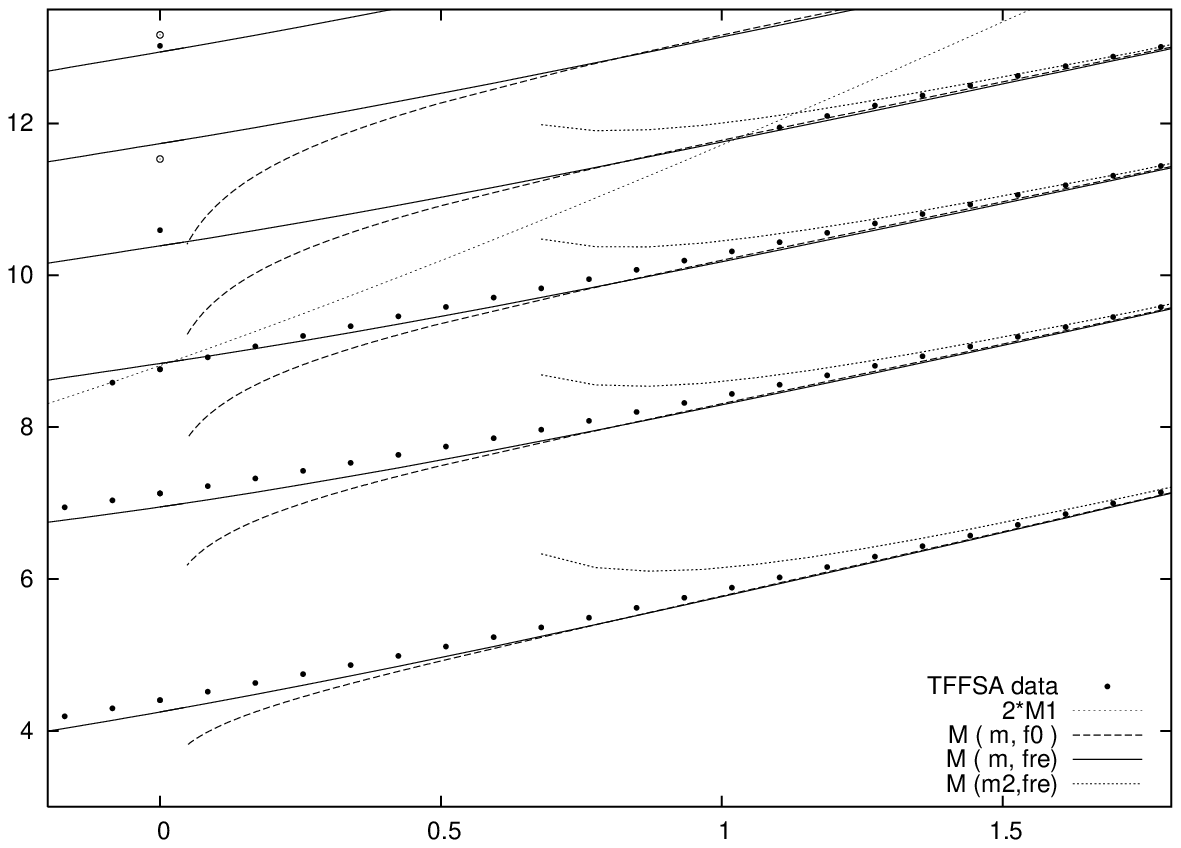}
\caption{\small Plots of few lowest meson masses (in the units of
$|h|^{16\over 15}$), vs the parameter $\eta$. The bullets
  $\bullet$ represent the actual masses (exact values at $\eta=0$, and
  the TFFSA data elsewhere. The uncertainties in the latter data are
  smaller then the size of the bullets). The fine dotted line is the
  stability threshold $2 M_1$. The dashed lines are the masses 
${\widetilde M}_n (m,f_0)$ from the BS equation with no
  renormalization effects taken into account. The solid lines
  are plots of ${\widetilde M}_n (m, f_{\rm re})$. Effect of the first
correction term in \eqref{mqseries} is shown by the dotted
lines. The circles $\circ$ indicate position of the higher thresholds 
$M_1 + M_2$ and $M_1 + M_3$ at $\eta=0$.}
\label{fig-deltam2}
\end{figure}
roughly the same region where the higher-order corrections to the
string tension $f$ become significant, see Fig.6. The dramatic drop of 
the BS masses ${\widetilde M}_{n}(m,f_0)$ at $\eta\to 0$ visible in the plot 
is certainly related to the singular behavior of $f_0$ in this
limit. Therefore it is reasonable to expect that replacing $f_0$ with 
renormalized string tension $f$ would take care of the most part of
the mismatch at small $\eta$. As was already said, it is not
completely clear how to determine the string tension $f$ beyond
the perturbation theory, but it is unlikely we can be off too far if
we take, say, the real part of the difference \eqref{deltaf}, i.e. set 
\bea\label{fre}
f = f_{\rm re} \equiv \Re e\,\big({\cal F}_{\rm meta} - {\cal F}_{\rm
  vac}\big).
\eea
In particular, we recall that $f_{\rm re}$ has finite limit at
$\eta=0$, and in fact analytic around this point. As the result, the
BS masses ${\widetilde M}_{n}(m,f_{\rm re})$ are also analytic at $\eta=0$,
and their expansions around this point follow from \eqref{yseries}
(where this time one sets $\lambda=f_{\rm re}/m^2$), and
\eqref{deltafser}. We show some coefficients of these expansions in
Table 3, to compare with the coefficients in \eqref{mnseries}. To get 
further feeling of the matter, we have computed numerically few
lowest BS masses ${\widetilde M}_n$ using $f_{\rm re}$ instead of
$f_0$ in \eqref{bs}; the results (also shown in Fig.8) indeed come
rather close to the actual masses at all positive $\eta$. It is
tempting to attribute the remaining mismatch of the lowest 
BS masses ${\widetilde M}_{n}(m,f_{\rm re})$ (clearly visible in Fig. 8 
for $n=1,2,3$) to the effect of the quark mass renormalization, so
far disregarded. Indeed, if the correction to the quark mass is
positive (as the $\lambda^2$ term in \eqref{mqseries} is), its effect
would be to shift the meson masses upward. Unfortunately, nothing is 
known about the higher-order corrections to $m_q$ (other then the term
\eqref{mqtwo}, that is), let alone anything beyond the perturbative 
expansion. Just adding the known $\lambda^2$ term, while noticeably 
improving the BS masses at sufficiently large $\eta$ (as was seen in
Fig.7), leads to substantial overestimate
of the actual masses at small $\eta$ (Fig.8). This suggests that the 
properly defined
$m_q$ remains close to the bare quark mass $m$ at all positive $\eta$,
being perhaps no greater then $(0.1)\,\,|h|^{8/15}$ at $\eta=0$. 

Since the two-quark approximation can not account for the possible
instability of the heavier mesons with respect to decays into the
lighter ones, the BS masses ${\widetilde M}_n$ remain real at all positive
$\eta$. This is unlike the actual mesons which turn to resonances as
soon as their masses cross the stability threshold $2 M_1$. In fact
the behavior of the actual masses $M_n (\eta)$ near the threshold is
rather subtle (see footnote$^6$ on page 4 for comment); this complex 
threshold behavior of the masses $M_3 (\eta)$ and $M_4 (\eta)$ is even
visible in Fig.8. Of course this subtleties are not captured by the BS
equation \eqref{bs} - the BS masses ${\widetilde M}_n$ go right through,
without any anomalies at the threshold. This deficiency is the reason
why in the Table 3, while ${\widetilde M}_{3}^{(0)}$ is rather 
close to $M_{3}^{(0)}$, the slope ${\widetilde M}_{3}^{(1)}$ and the
curvature ${\widetilde M}_{3}^{(2)}$ are so much off - the point
$\eta=0$ is too close to the stability threshold $\eta_3 = -0.136(1)$
of the third particle. Let us recall that it is when the meson mass
approaches the threshold that the multiquark corrections to the
``regular'' part of the BS kernel, Eq.\eqref{newker}, are expected to 
play a prominent role, and it is plausible that including such
multiquark corrections would help to understand better the threshold 
behavior. 

In this paper we do not discuss the resonance part of the spectrum,
putting this task away for another time. However let us make few remarks
here. Although the BS equation \eqref{bs} can not account for the
instability of the mesons above the threshold (and thus disregards the
imaginary parts of their masses), it likely captures the most of the
real parts of the resonance masses. This is not inconsistent with
the pattern readily observed in Fig.8. For example, the BS masses 
${\widetilde M}_4 (m, f_{\rm re})$ and ${\widetilde M}_6 (m, f_{\rm
  re})$ at $\eta \to 0$ come reasonably close to the exact masses of
the fourth and the fifth of the stable particles of the
$\eta=0$ theory, suggesting that analytic continuations of the 
actual masses $M_4 (\eta)$ and $M_6 (\eta)$ to the point $\eta=0$ 
reproduce those exact masses. This rises a question about the
fate of the mass $M_5 (\eta)$ at $\eta\to 0$. Now, observe that the BS 
mass $M_5 (m,f_{\rm re})$ at $\eta\to 0$ lands far from $M_{5}^{(0)}$
(this is cleaarly seen in the last line of Table 3) but rather close 
to $M_{1}^{(0)} +M_{2}^{(0)}$, the sum of two lowest masses of the 
$\eta=0$ theory. The most economic scenario then implies that 
for the actual masses we have $M_5 (0)-M_{1}(0) - M_2 (0) =0$, and
at $\eta<<1$ the state with mass $M_5(\eta)$ can be interpreted as
a weakly coupled bound state of the two particles $M_1$ and $M_2$.
Of course it would be interesting to elaborate quantitative details of
such scenario (and its versions for the higher resonances); again, we
hope to address this problem in the future.

\section{Discussion}

In this paper we refine technique which allows for quantitative
understanding of the mass spectrum of the IFT, in the low-T domain, in 
terms of ``mesons'' consisting predominantly of two quarks bound by a
confining interaction. Thus we develop further the idea originally due to Wu
and McCoy \cite{mccoy2}. Systematic approach based on the idea of the mesons
being essentially the two-quark constructs leads to the Bethe-Salpeter
equation, Eq.\eqref{bs}, yielding a discrete spectrum of the meson
masses. The Bethe-Salpeter equation is itself an approximation since
it essentially ignores multi-quark components of the bound states. The
approximation becomes exact in the weak field limit $\eta\to\infty$,
but we show that it reproduces the masses of stable particles with
reasonable accuracy at all positive $\eta$. 

There is much room for further development. First, it seems to be
possible (albeit involved) to take into account the multi-quark
corrections perturbatively. This problem was briefly discussed in
Section 6. It would be especially important to understand the role
of the multiquark corrections to the ``regular'' part of the BS
kernel, the term $\Delta {\mathbb G}_{P}^{({\rm reg})}$ in
Eq.\eqref{newker}. That would help to get at least some insight into
the  way the meson masses behave near the stability
threshold. Understanding of the above-threshold (i.e. resonance) part
of the spectrum is another open problem (see however \cite{rut}).

As was mentioned in the Introduction, the main motivation of this work
originated from our attempts to understand analytic properties of the
IFT masses $M_n (\eta)$ as the functions of complex $\eta$. There are
reasons to believe that at least some singularities of these functions
are mimicked by the BS masses ${\widetilde M}_n (m,f)$, considered as the
functions of complex variable ${\bar\eta} = m/\sqrt{f}$, and
understanding analytic properties of the BS masses is a much more
manageable problem. For instance, one type of singularities -
those appearing due to collisions of the roots of the equation
\eqref{nueq} - was mentioned at the end of Sect.7. These are
square-root branching points, located somewhere at $|{\rm arg}\,
{\bar\eta}| > 3\pi/4$ (We plan to discuss these singularities in
more details elsewhere). It is likely that the actual IFT masses 
$M_n (\eta)$ have
similar singularities at $|{\rm arg}\,\eta| > 11\pi/15$. Another
question of much interest concerns locations of possible zeros of the
BS masses ${\widetilde M}_n$ on the Riemann surface of ${\bar\eta}$. 
At least some of such zeros may signal divergence of the
correlation length in the full-fledged IFT, where one would expect to 
observe associated critical points. In this manner, vanishing of the
lowest meson mass $M_1 (\eta)$ at $\eta=(2.4295..)\,e^{\pm\,i{11\over
    15}\pi}$ is related to the Yang-Lee edge singularity \cite{fz1}.
It is certainly interesting to contemplate the possibility that the 
Yang-Lee edge singularity is but one of many critical points one can
bump into when wandering on the Riemann surface of $\eta$ in the IFT.
To some extent, the present paper can be regarded as a preparation to
this analysis, and we hope to address this family of problems in 
future work.

Finally, some of the techniques developed here, in particular the weak
coupling expansions in Sect.5, can be useful in other 2D theories
which exhibit confinement of quarks, notably the 2D model of
multicolor QCD (the 't Hooft model). More generally, there is a large 
class of systems having degeneracy of the ground state, which are
integrable (with factorizable S-matrix of kinks), but not
free. Typical perturbations of such theories lift the ground state
degeneracy and creates confining force for the kinks. It seems
important to extend the approach developed here to such systems.

\section*{Appendix}

\appendix

\section{Infinite-momentum limit of Eq.\eqref{inteq}}

Here we derive the $P\to \infty$ limit of the equation
\eqref{inteq}. We assume that $\varepsilon(p)$ has exact asymptotic 
behavior \eqref{epsas}, and that $\Delta E(P)$ has the form
\eqref{deltae}, in particular
\bea\label{deleass}
\Delta E = |P\,| + {{M^2}\over{2\,|P\,|}} + O(|P\,|^{-3})\ .
\eea
Both sides of \eqref{inteq} can be expanded
in inverse powers of $P$. Assume that $\Psi_P(p)$ is normalized in
such a way that at $|p| < P/2$ it has finite nonzero limit 
as $P\to +\infty$. Under such normalization, $\Psi_P(p)$ with
$|p|>P/2$ tends to zero as $P\to +\infty$. More precisely, if $u = 
2p/P$ is fixed,
\bea\label{psipass}
\Psi_P (p) = \bigg\{ {{O\,\,(\ 1\ ) \ \quad {\rm at}\quad |u\,|<1}\atop
{O(P^{-2}) \quad {\rm at}\quad |u\,|>1}}\bigg\} \quad {\rm as}\quad
P\to \infty\,.
\eea
Indeed, consider the left-hand side of the equation \eqref{inteq}. The
coefficient in front of $\Psi_P (p)$ has different $P\to\infty$
behavior depending on whether $|u|$ is smaller or greater then $1$,
namely 
\bea\label{lhsass}
{\rm l.h.s.\ of\ (3.11)} \to 
\bigg\{{{{P^{-1}}\,\big[{{2 m^2}\over{1-u^2}}-{{M^2}\over
    2}\big]\,\Phi(u) \quad \ {\rm at}\quad
    |\,u\,|<1}\atop{\ \ P\ \ \big[\,|\,u\,| - 1\,\big]\ \ \Phi(u) \qquad {\rm
      at} \quad |\,u\,|>1}}\bigg\} \quad {\rm as}\quad P\to\infty\,,
\eea
where
\bea\label{psiu}
\Phi( u )= \lim_{P\to\infty}\,\Psi_P \,({{uP}/2})\ .
\eea
On the other hand, the kernel $G_P(p|q)$ decays as $1/P^2$ at all 
$\ p,\,q \sim P$, and hence the right-hand side of \eqref{inteq}
is $\sim P^{-1}$ at all $u$. As the consequence, the limiting
wave-function $\Phi(u)$, Eq.\eqref{psiu}, must vanish outside the
interval $|u|<1$,
\bea\label{psiout}
\Phi(u) = 0 \quad {\rm at} \quad |\,u\,|>1\ .
\eea
Comparing the $1/P$ terms in both sides of \eqref{inteq}, we obtain
the equation
\bea\label{bsu}
\bigg[{{m^2}\over{1-u^2}} - {{M^2}\over {4}}\bigg]\Phi(u) =
{f_0}\ \dashint_{-1}^{1}\,F(u|v)\,\,\Phi(v)\,\,{{dv}\over{2\pi}}\ ,
\eea
where
\bea\label{uvkernel}
F(u|v) =
{1\over{\sqrt{(1-u^2)(1-v^2)}}}\,
\bigg[{{1-uv}\over{(u-v)^2}}-{{1+uv}\over{(u+v)^2}}+{{uv}\over
    4}\bigg]\ .
\eea
Changing variables $u=\tanh\theta$, $v = \tanh\theta'$ one arrives at
\eqref{bs} with $\Psi(\theta) = \Phi(\tanh\theta)$.

Note that the equation \eqref{bsu} is very similar to the
Bethe-Salpeter equation determining the mass spectrum of mesons in the
't Hooft's model \cite{thooft}. When the two quarks have
equal masses the latter differs from \eqref{bsu} only in the detailed
form of the kernel, which in that case is somewhat simpler, 
\bea\label{thooftker}
F_{\rm \ 't\,Hooft}(u|v) = {1\over{(u-v)^2}} 
\eea
(also, the parameter $f_0$ is to be interpreted as the square of
the gauge coupling constant).

\section{Derivation of the Eq.\eqref{deltaint}}

Here we evaluate the action of the operator ${\hat H}-M^2$ (with $\hat
H$ defined in \eqref{hhat}) on the wavefunction \eqref{psioo}. It is
useful to split the letter into two pieces,
\bea\label{formsa}
\Psi_0 (\theta) = \Psi_{+}(\theta) - \Psi_{+}(-\theta)\ ,
\eea
with
\bea\label{psiplus}
\Psi_{+}(\theta) =
{1\over 2}\,\int_{-\infty}^{\infty}\,
{{e^{{i\over\lambda}S(\beta)}}\over{\sinh(\theta+\beta-i0)}}\,d\beta\,,
\eea
and evaluate the corresponding two terms separately. In fact we need
only to evaluate $\big[{\hat H}-M^2\big]\Psi_{+}(\theta)$; the desired
result is then obtained by taking the part antisymmetric with respect
to the reflection $\theta\to -\theta$. 
It will be also convenient to represent the operator ${\hat H}-M^2$
as the sum of two terms
\bea\label{splitt}
[{\hat H}-M^2]\Psi(\theta) = 4{m^2}\,\cosh^2\theta\,\,
\big[\,{\hat \Omega}\Psi(\theta) + {\hat G}\Psi(\theta)\,\big]\ ,
\eea  
where
\bea\label{bigomega}
{\hat \Omega}\Psi(\theta) = 
\Omega(\theta)\Psi(\theta)\ , \quad {\rm with} \quad
\Omega(\theta) = - {\partial\over{\partial\theta}} S(\theta) =  
1 - {{M^2}\over{4 m^2}}\,{1\over{\cosh^2\theta}} \ ,
\eea
and 
\bea\label{ghat}
{\hat G}\Psi(\theta) = - \lambda\,\dashint_{-\infty}^{\infty}\,
G_0 (\theta-\theta')\,
\Psi(\theta)\,{{d\theta'}\over{2\pi}} + {\lambda\over
  4}\,{{\sinh\theta}\over{\cosh^2\theta}}\,{\bar\Psi}\ .
\eea
In the last equation we have introduced the notations
\bea\label{gopsibar}
G_0 (\theta-\theta') =
2\,{{\cosh(\theta-\theta')}\over{\sinh^2(\theta-\theta')}}\ , \qquad
{\bar\Psi} =
\int_{-\infty}^{\infty}\,{{\sinh\theta}\over{\cosh^2\theta}}\,
\Psi(\theta)\,{{d\theta}\over{2\pi}}\ .
\eea

Let us first transform the integral
\bea\label{transa}
-\lambda\,\dashint_{-\infty}^{\infty}\,G_0
(\theta-\theta')\Psi_{+}(\theta')\,{{d\theta'}\over{2\pi}} =
&&\\
{d\over{d\theta}}\,\,\dashint_{-\infty}^{\infty}\,
{\lambda\over{\sinh(\theta-\theta')}}\,{{d\theta'}\over{2\pi}}&&
\int_{-\infty}^{\infty}\,{{e^{{i\over\lambda}S(\beta)}\,\,d\beta}
\over{\sinh(\theta'-\beta-i0)}}\ . \nonumber
\eea
The order of integrations can be interchanged, and at fixed real
$\beta$ the contour of integration over $\theta'$ can be shifted
downward, $\theta' =\alpha -i\pi/2$; this leaves behind half of
the residue at the pole $\theta' = \theta$,
\bea
(B.8) = {i\over 2}\,\int_{-\infty}^{\infty}\,
{\partial\over{\partial\theta}}\,
{{\lambda\,\,\,e^{{i\over\lambda}S(\beta)}\,d\beta}\over
{\sinh(\theta+\beta-i0)}} +
{d\over{d\theta}}\,\int_{-\infty}^{\infty}\,{{d\beta}\over{2\pi}}\,
\int_{-\infty}^{\infty}\,
{{\lambda\,\,\,e^{{i\over\lambda}S(\beta)}\,d\alpha}
\over{\cosh(\theta-\alpha)\cosh(\alpha + \beta)}}\ . \nonumber
\eea
In the first term we transfer the derivation over $\theta$ to the
integration variable $\beta$, and then integrate by parts. The
integral over $\alpha$ in the second term is evaluated in closed
form. One finds 
\bea\label{transb}
(B.8) =
-{1\over 2}\,\int_{-\infty}^{\infty}\,
{{\Omega(\beta)\,e^{{i\over\lambda}S(\beta)}\,d\beta} 
\over{\sinh(\theta+\beta-i0)}} + {\lambda\over\pi}\,\,
\int_{-\infty}^{\infty}
\,\bigg[{\partial\over{\partial\beta}}\,
{{\theta+\beta}\over{\sinh(\theta+\beta)}}\bigg]\,\,
e^{{i\over\lambda}S(\beta)}\,\,d\beta\ ,
\eea
where the factor $-\Omega(\beta)$ in the first term came from the
derivative of $S(\beta)$, see Eq.\eqref{bigomega}. Note that when
plugged into \eqref{splitt}, the first term combines nicely with 
the term $\Omega(\theta)\Psi_{+}(\theta)$, to produce
\bea\label{transc}
\int_{-\infty}^{\infty}\,\,{1\over 2}\,
{{\Omega(\theta)-\Omega(\beta)}\over
{\sinh(\theta+\beta-i0)}}\,\,e^{{i\over\lambda}S(\beta)}\,\,d\beta\,.
\eea
Since the numerator in the integrand vanishes at $\beta=-\theta$,
this point is no longer singular, and the shift $-i0$ in the
denominator is irrelevant. In fact, the integrand in \eqref{transc}
can be reduced to
\bea\label{omegas} 
{{M^2}\over{8 m^2}}\,\,
{{\sinh(\theta-\beta)}\over{\cosh^2\theta\,\,\cosh^2\beta}}\,\,
e^{{i\over\lambda}S(\beta)}\,.
\eea

Next, when applied to $\Psi_{+}$, the second term in right-hand side
of \eqref{ghat} involves the factor
\bea\label{psiplusbar}
{\bar\Psi}_{+} = {1\over 2}\,\int_{-\infty}^{\infty}\,
{{\sinh\theta}\over{\cosh^2 \theta}}\,\,{{d\theta}\over{2\pi}}\,\,
\int_{-\infty}^{\infty}\,{{e^{{i\over\lambda}S(\beta)}}\over
{\sinh(\theta+\beta-i0)}}\,d\beta && = \nonumber\\
\int_{-\infty}^{\infty}\,\,\,\,\bigg[{i\over
    4}\,\,{{\sinh\beta}\over{\cosh^2\beta}} -
    {1\over{2\pi}}\,\,{\partial\over{\partial\beta}}
{\beta\over{\cosh\beta}}\bigg]&&e^{{i\over\lambda}S(\beta)}\,\,d\beta\,,
\eea
where the expression in the second line results from explicit
evaluation of the integral over $\theta$. 

Putting the Eq's. \eqref{bigomega}, \eqref{ghat}, \eqref{transb}, and
\eqref{psiplusbar} together, we find
\bea\label{transd}
\big[{\hat\Omega}+{\hat G}\big]\Psi_{+}(\theta) =
\int_{-\infty}^{\infty} \,\,
\bigg[{{M^2}\over{8 m^2}}\,\,{{\sinh(\theta-\beta)}\over{\cosh^2\theta\,\,\cosh^2\beta}} +
  \qquad\qquad\qquad\qquad\qquad\nonumber\\
{\lambda\over\pi}\,{\partial\over{\partial\beta}}
\bigg({{\theta+\beta}\over{\sinh{(\theta+\beta)}}}-{1\over{8}}\,\,
{{\sinh\theta}\over{\cosh^2\theta}}\,\,{{i\pi/2
    +\beta}\over{\cosh\beta}}\bigg)\bigg]\,\,
e^{{i\over\lambda}S(\beta)}\,\,d\beta\,.
\eea
Finally, isolating the odd part of this expression, one arrives at 
\eqref{deltaint}.

\section*{Acknowledgments}

One of us (AZ) is pleased to acknowledge kind hospitality extended to him
at the Institute of Physics of the University of Bonn, and at the Ecole
Normale Sup\'erieure, where parts of this work was done. Supports from
the Alexander von Humboldt Foundation, and Blaise Pascal Research
Chair, which made these visits possible, are gratefully
appreciated. We would like to express gratitude to Alyocha
Zamolodchikov, Sergei Lukyanov, and Volodya Fateev for many
stimulating discussions and helpful remarks. AZ also benefited a lot 
from conversations with Vladimir Bazhanov, Vladimir Kazakov, and
Vladimir Rittenberg. Research of AZ is supported by DOE grant \#
DE-FG02-96ER40949.


\begin{thebibliography}{99}

\bibitem{drouffe}{C.\ Itzykson, J.--M.\ Drouffe,
{\it Statistical field theory},
Cambridge University Press, 1989.}

\bibitem{cft}{P. Di Francesco, P. Mathieu, D. S\'en\'echal, {\it
  Conformal Field Theory}, Springer, 1996}

\bibitem{mccoy2} B.M. McCoy, T.T. Wu, {\it
Two-dimensional Ising field theory in a magnetic field: breakup of the
cut in the two-point function}, Phys.\ Rev. D 18 (1978) 1259--1267.

\bibitem{fz1} P. Fonseca, A. Zamolodchikov,
{\it Ising field theory in a
magnetic field: analytic properties of the free energy}, J.\ Stat.\ Phys.
110 (2003) 527--590,\  hep-th/0112167

\bibitem{me1} A.B. Zamolodchikov {\it Integrals of motion and S
  matrix of the (scaled) $T = T_c$ Ising model with magnetic field.}, 
Int.J.Mod.Phys. A4 (1989) 4235,  {\it Integrable field theory from
  conformal field theory}, Advanced Studies in Pure Mathematics 19
  (1989) 641-674

\bibitem{delfino2} G. Delfino, G. Mussardo, P. Simonetti,
{\it Non-integrable
quantum field theories as perturbations of certain integrable models},
Nucl. Phys. B 473 (1996) 469--508;~{hep-th/9603011}.

\bibitem{mccoy1} T.T. Wu, B.M. McCoy, {\it The two-dimensional Ising
model}, Harvard University Press,\ 1973.

\bibitem{yl} C.N. Yang, T.D. Lee, {\it Statistical Theory of Equation of
  State and Phase Transitions. I. Theory of Condensation}, Phys.Rev. 87
  (1952) 404-409; T.D. Lee, C.N. Yang, {\it Statistical Theory of Equation of
  State and Phase Transitions. II. Lattice Gas and Ising Model}, Phys.Rev. 87
  (1952) 410-419. 

\bibitem{fisher} M.E. Fisher, {\it Yang-Lee Edge Singularity and
  $\varphi^3$ Field Theory}, Phys.Rev.Lett. 40, (1978) 1610-1613.

\bibitem{cardy1} J.L. Cardy, {\it Conformal Invariance and the
  Yang-Lee Edge Singularity in Two Dimensions}, Phys.Rev.Lett. 54
  (1985) 1354-1356. 

\bibitem{cardy2} J.L. Cardy, G. Mussardo, {\it S-Matrix of the
  Yang-Lee Edge Singularity in Two Dimensions}, Phys.Lett. B225 (1989)
  275-278.

\bibitem{tcsa} V.P. Yurov, Al.B. Zamolodchikov, {\it Truncated
  Conformal Space Approach to Scaling Lee Yang Model},
  Int.J.Mod.Phys. A5 (1990) 3221-3245; V.P. Yurov,
  Al.B. Zamolodchikov, {\it Truncated Fermionic Space Approach to the
  Critical 2D Ising Model with Magnetic Field}, Int.J.Mod.Phys. A6
  (1991) 4557-4578.

\bibitem{rut} S.B. Rutkevich,
{\it Large-N excitations in the ferromagnetic Ising
  field  theory in a small magnetic field: mass spectrum and decay
  widths}, Phys.Rev.Lett. 95 (2005) 250601; hep-th/0509149

\bibitem{thooft} G. 't Hooft, {\it A two-dimensional model for mesons},
Nucl.Phys. B~75 (1974) 461--470

\bibitem{landau} L.D. Landau and E.M. Lifshitz, {\it Quantum Mechanics
  (Non-relativistic Theory)}, Pergamon Press, 1991

\bibitem{res1} G. Delfino, P. Grinza, G. Mussardo,
{\it Decay of particles above threshold in the Ising field theory
  with magnetic field}, Nucl.Phys. B737 (2006) 291-303; hep-th/0507133

\bibitem{tracy} T.T. Wu, B.M. McCoy, C.A. Tracy, E. Barouch, {\it
Spin-spin correlation functions for the two-dimensional Ising model:
exact theory in the scaling region}, Phys. Rev. B~13 (1976) 316--374.

\bibitem{karowski} B. Berg, M. Karowski, P. Weisz, {\it Construction
  Of Green Functions From An Exact S Matrix}, Phys.Rev. D19 (1979).

\bibitem{qft} C. Itzykson, J.-B. Zuber, {\it Quantum Field Theory},
  McGrow-Hill, 1980.


\bibitem{bars} I.Bars, M.B. Green, {\it Poincar\'e- and gauge
-invariant two-dimensional quantum chromodynamics}, Phys. Rev. D~17,
  (1978), 537--545

\bibitem{mccoy3} B.M. McCoy, T.T. Wu, {\it Two-dimensional Ising model
  near $T_c$: Approximation for small magnetic field}, Phys.Rev. B18
  (1978) 4886-4901.

\bibitem{fz2} P. Fonseca, A. Zamolodchikov, {\it Ward identities and
integrable  differential equations in the Ising field theory},
RUNHETC-2003-28, Sep 2003;\ hep-th/0309228

\bibitem{nucleation} A.F. Andreev, {\it Singularity of Thermodynamic
  Quantities at a First Order Phase Transition Point}, JETP 18 (1964)
  1415-1416; M.E. Fisher, University of Colorado Summer School Lectures,
  Boulder, 1964; J.S. Langer, {\it Theory of Condensation Point},
  Ann.Phys.41 (1967) 108-157.

\bibitem{volosha} Voloshin, I.Yu. Kobzarev, L.B. Okun, M.B. Voloshin,
  {\it Bubbles in metastable vacuum}, Sov.J.Nucl.Phys. 20 (1975)
  644-646;  S. Coleman, {\it Fate of the false vacuum: Semiclassical
  theory}, Phys.Rev. D15 (1977) 2929-2936.

\bibitem{matthias} W. Krauth, M. Staudacher, {\it Nonintegrability of
  two-dimensional QCD}, Phys.Lett. B~388 (1996) 808--812; \
  hep-th/9608122

\bibitem{fatey} V.A. Fateev, {\it The exact relations between the
  coupling constants and the masses of particles for the integrable
  perturbed Conformal Field Theories}, Phys.Lett. B 324 (1994) 45-51.

\bibitem{slopes} P. Grinza, A. Rago, {\it Study of the 2-D Ising model
  with mixed perturbation},
Nucl.Phys.B651 (2003), 387-412; hep-th/0208016

\bibitem{callan} C.G. Callan, N. Coote, D.J. Gross, {\it
  Two-dimensional Yang-Mills theory: A model of quark confinement},
  Phys.Rev. D13 (1976) 1649-1669.

\bibitem{delfino1} G. Delfino, {\it Integrable field theory and
  critical phenomena: The Ising model in a magnetic field},
  J.Phys. A37 (2004) R45. 

\bibitem{caselle} M. Caselle, P.Grinza, A. Rago, {\it Amplitude ratios
  for the mass
  spectrum of the 2d Ising model in the high-$T$, $H \neq 0$ phase},
  J.Stat.Mech. 0410 (2004), P009, e-Print Archive: hep-lat/0408044.


\end{thebibliography}
\end{document}